\documentclass[acmtog, nonacm,s balance=false, prologue, table]{acmart}
\usepackage{amsmath}
\usepackage{booktabs} 
\usepackage{float}
\usepackage{layouts}
\usepackage[ruled,vlined]{algorithm2e}
\usepackage{bbm}
\usepackage{wrapfig}

\usepackage{appendix}

\definecolor{forestgreen}{rgb}{0.13, 0.55, 0.13}
\definecolor{lava}{rgb}{0.81, 0.06, 0.13}
\definecolor{magenta}{rgb}{0.7, 0.0, 1.0}
\definecolor{softblue}{rgb}{0.0, 0.5, 1.0}

\newcommand{\ty}[1]{\textbf{\textcolor{softblue}{}}}
\newcommand{\danny}[1]{\textbf{\textcolor{orange}{}}}
\newcommand{\dave}[1]{\textbf{\textcolor{magenta}{}}}
\newcommand{\todo}[1]{\textbf{\textcolor{purple}{}}}

\definecolor{staticColor}{HTML}{005AB5}

\definecolor{dynamicColor}{HTML}{DC3220}

\newcommand{\reffig}[1]{Fig.~\ref{fig:#1}}
\newcommand{\refeq}[1]{Eq.~(\ref{eq:#1})}
\newcommand{\refsec}[1]{Sec.~\ref{sec:#1}}

\usepackage{amsmath}

\DeclareMathOperator*{\argmin}{argmin}

\usepackage{mfirstuc}
\MFUnocap{et}
\MFUnocap{al}

\newcommand{\vc}[1]{\mathbf{#1}}

\newcommand{\R}{\mathbb{R}}

\newcommand{\numverts}{{|\mathcal{V}|}}

\begin{document}

\title{Sparse, Geometry- and Material-Aware Bases for Multilevel Elastodynamic Simulation}

\author{Ty Trusty }
\email{trusty@cs.toronto.edu}
\affiliation{
\institution{University of Toronto}
\country{Canada}
}
\affiliation{
\institution{Adobe}
\country{USA}
}
\author{David I.W. Levin}
\email{diwlevin@cs.toronto.edu}
\affiliation{
 \institution{University of Toronto}
 \country{Canada}
}
\affiliation{
 \institution{NVIDIA}
 \country{Canada}
}
\author{Danny M. Kaufman}
\email{dannykaufman@gmail.com}
\affiliation{
\institution{Adobe}
\country{USA}
}


%

\begin{abstract}
We present a multi-level elastodynamics timestep solver for accelerating incremental potential contact\ \cite{IPC} (IPC) simulations. Our method retains the robustness of gold standard IPC in the face of intricate geometry, complex heterogeneous material distributions and high resolution input data without sacrificing visual fidelity (per-timestep relative displacement error of $\approx1\%$). The success of our method is enabled by a novel, sparse, geometry- and material-aware basis construction method which allows for the use of fast preconditioned conjugate gradient solvers (in place of a sparse direct solver), but without suffering convergence issues due to stiff or heterogeneous materials. The end result is a solver that produces results visually indistinguishable and quantitatively very close to gold-standard IPC methods but up to $13\times$ faster on identical hardware.
\end{abstract}

\begin{teaserfigure}
  \includegraphics[width=\textwidth]{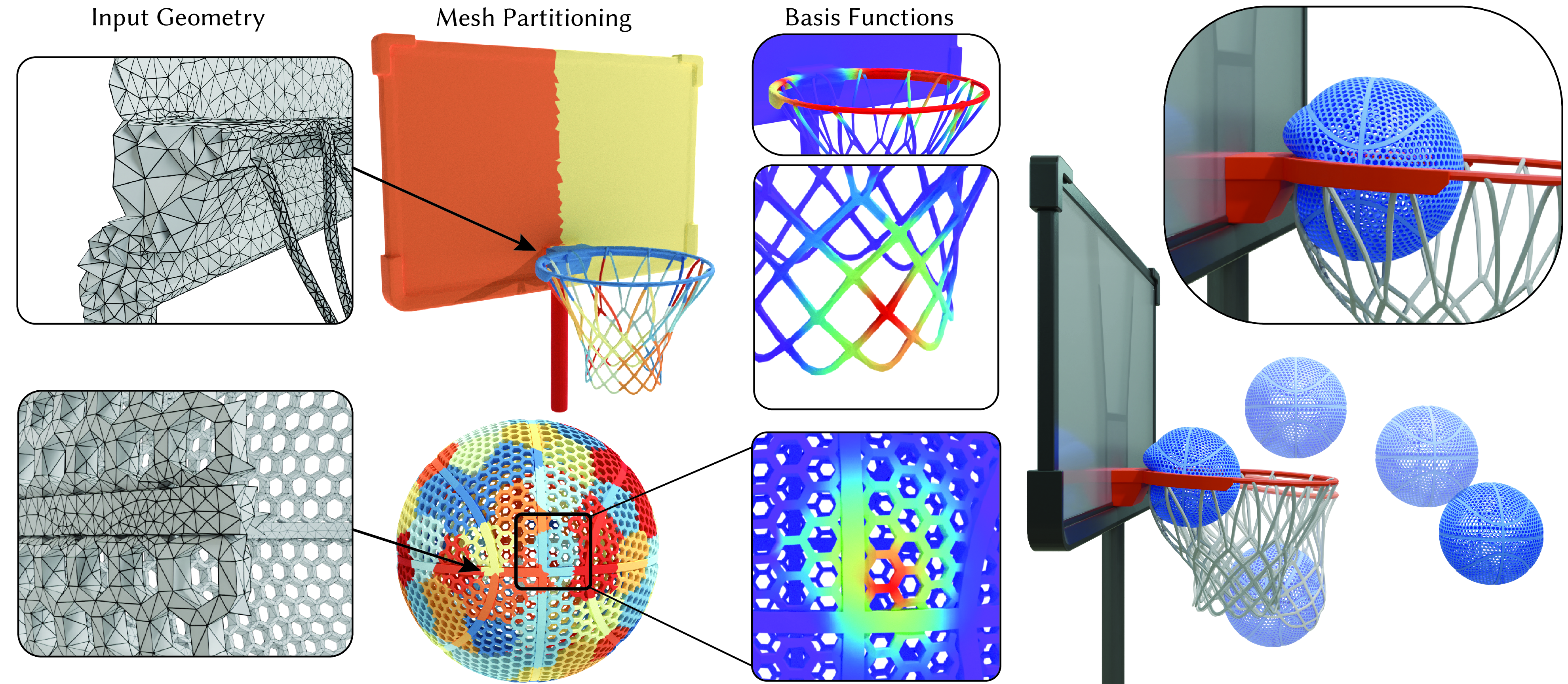}
  \caption{They shoot, they score! We simulate an airless, lattice-design basketball as it swishes through a steel basketball hoop using our new multi-level elastodynamics solver and novel sparse basis construction with a $9.8\times$ speedup over a comparable IPC simulation. Every object in this scene is represented by a high-resolution volumetric tetrahedralization (220K Tetrahedra), highlighting our method's ability to simulate highly intricate geometry and heterogeneous materials at large-scale.}
  \label{fig:teaser}
\end{teaserfigure}

\maketitle


\section{Introduction}
Recent work in robust frictional contact solvers for elastodynamics has enabled breakthroughs in the complexity of scenes simulated in computer graphics, robotics and mechanical engineering. The modern approach to these simulations is to leverage barrier potentials for contact, and mollified frictional pseudopotentials, to enable effective and stable implicit integration of geometrically and materially complex mechanical systems. Algorithms based on these building blocks have been developed for volumetric and co-dimensional solids along with rigid and quasi-rigid bodies. 

Unfortunately, robustness comes at a computational cost. The gold-standard, incremental potential contact (IPC) method\ \cite{IPC},  relies heavily on Newton-type optimization, and can suffer from ill-conditioned Hessians caused by stiff contact forces or strongly heterogeneous material distributions. This ill-conditioning, in turn, often requires the use of a sparse direct solvers in the inner optimization loops. Such solvers are robust, but do not scale effectively to parallel architectures. Further, even with a robust solver in-hand, the stiff and ill-conditioned nature of the IPC optimization problem can significantly slow overall convergence. The often hundreds-of-thousands or millions of tetrahedra required to represent complex geometries or material distributions then only exacerbates these issues. As simple performance acceleration solutions, such as early solver terminations, can and will lead to excessive damping, trajectory errors, and other unacceptable artifacts, this motivates the search for alternative avenues for improving performance. 

There are many fast elastodynamics algorithms in the literature. While their low-level design details vary, a typical approach is to directly tackle the performance of the optimization loop itself via application of iterative solvers (e.g., via Krylov subspace methods such as Preconditioned Conjugate Gradient (PCG), local-global methods such as ADMM\ \cite{overby2017admm}, or more elaborate local-stencil methods such as Vertex Block Descent\ \cite{chen2024vertex}). These methods demonstrate impressively fast and large-scale results. However, what they gain in speed, they sacrifice in terms of robustness. Krylov subspace methods require careful preconditioning and, even then, performance can and will significantly degrade when large variations in object stiffnesses are present. Local-stencil, and local-global approaches limit how far and fast deformation can propagate across the simulation mesh. Stiff objects and stiff inclusions discretized at high resolution both cause stagnating convergence, and require per-example careful tuning of algorithm parameters, including iteration counts. This coupling of geometric representation, material distribution, and algorithm parameters often forces per-example parameter tuning which makes these methods difficult to use as opaque box solvers at large scale, and can even prohibit simulations altogether. 

In this work, we take a similar approach, but propose different criteria, attempting to maximally accelerate frictionally contacting elastodynamic simulation without removing the robustness and generality guarantees of previous work. To this end, we start with the traditional IPC implicit integration scheme and introduce a new three-level timestep solver, which is largely immune to the issues discussed above. The key to our approach is a novel \emph{geometry and material aware} mid-level subspace layer which automatically provides an intermediate representation well-suited to input volumetric physical objects, including those with challenging material and geometric heterogeneities, and enables the use of iterative linear solvers throughout our solver. After computing a search direction in this subspace, we hand it off it to our final level, which, using full-space discretization, "polishes" our solution to recover high-frequency behaviors. 

Our subspace is theoretically sound, and practical to compute, with the primary computational operations being the solution of a set of standard PDEs (specifically the Laplace and biharmonic equations). The procedure is a fully automatic precomputation, requiring the user to simply choose a resolution for the subspace layer. As a result of our construction, our multilevel solver is then able to employ standard, preconditoned conjugate gradient solves at all three levels (for all simulation inputs) which parallelize well across multicore hardware. The result is an algorithm that is 7-13$\times$ faster than standard IPC, while exhibiting a per-timestep error of $1\%$ with respect to standard IPC across a number of highly complex geometries with complex material distributions.

\subsection{Contributions}
In summary our contributions are 
\begin{itemize}
    \item \emph{Sparse, Physics-Aware Basis Functions:} we introduce a new basis function generated via localized biharmonic solves that is compactly supported, geometry-aware, material-aware, and convergent.
    \item \emph{A Parallel Multilevel Subspace Solver:} we integrate our basis functions into a nonlinear three-level timestep solver composed of a global affine stage, local subspace solves, and a final full-space refinement, each designed to be compatible with block-diagonal preconditioned iterative linear solver methods.
    \item \emph{A Material- and Geometry-Aware Mesh Partitioning:} we introduce a new mesh partitioning custom-designed for our above basis functions, based on K-means clustering and stiffness-weighted heat geodesics.
    \item \emph{Subspace Integration via Basis-Aware Moment Fitting:} we accelerate our subspace Hessian assembly, a critical potential bottleneck, with a new extension of moment fitting to subspace bases.
\end{itemize}
\section{Related Work}
\label{sec:related}

Implicit time integration is the standard approach for elastodynamics simulation, and accordingly prior work has extensively explored all aspects: from speed, to accuracy, to robustness.
Our work focuses on an underexplored space, the construction of efficient elastodynamics solvers that retain the robustness and generality of their more computationally demanding counterparts.
Typical approaches to accelerating implicit solvers for elastodynamics can be  broadly categorized into two groups: (1) those that modify the discretization by using a coarser set of basis functions to reduce the number of degrees of freedom (DOFs), and (2) those that modify the nonlinear timestep optimization. Both approaches can also be used in conjunction, as we do in this work.
As we will later detail, our method follows a multilevel solver strategy that uses a reduced function space at the mid-level to help accelerate performance. 
Below we position our work within the context of both reduced-space methods and fast solvers for the elastodynamics problem.

\subsection{Basis Functions for Elastodynamics}
A key component of our method is the use of a reduced set of basis functions for medium-scale deformations. We assume access to a high-resolution base finite element discretization (here piece-wise linear elements), and our goal is to construct a coarser set of bases that approximates deformation in this full-space. The question is how to choose the basis functions with the most "bang for your buck".
%
Ideally, these basis functions should span the space of complex deformation, taking into account geometry, material distribution, and boundary conditions.
Next we review some potential candidates. 

\paragraph{Handle-Based Simulation}
Handle-based discretizations for simulation, introduced to graphics by \citet{gilles2011framebased}, enrich positions-based DOFs with high-order degrees of freedom.
With handle-based methods, the quality of the resulting bases is heavily dependent on weight (interpolating) function quality. Our work can be considered a generalization of \citet{faure2011sparsemeshless} which proposes sparse, material- and geometry- aware and basis functions based on compliance-weighted geodesics. They compute weights from each handle by constructing a geodesic-distance based on advancing a Voronoi diagram discretized on a voxel grid. The quality of output heavily depends on this discretization quality as a result. In our case, our weights are constructed implicitly from linear PDEs and the support boundary is extracted from the 
PDE solution, which avoids the time-consuming and sensitive explicit geometry processing of \citet{faure2011sparsemeshless}.

In the Finite Element context, these handle based methods can be interpreted as an instance of higher-order Finite Element Method (FEM). For instance, \citet{longva2020higherorder} uses high-order FE bases on an embedded coarse triangulation to achieve impressive results. As with handle-based methods, careful consideration is necessary when building this coarse triangulation to produce bases that conform to the underlying geometry and material distributions, and the method can not fully ameliorate all the issues with a regular embedding grid, such as ill-conditioning due to mostly empty cells.

\begin{figure}
    \includegraphics[width=\linewidth,keepaspectratio]{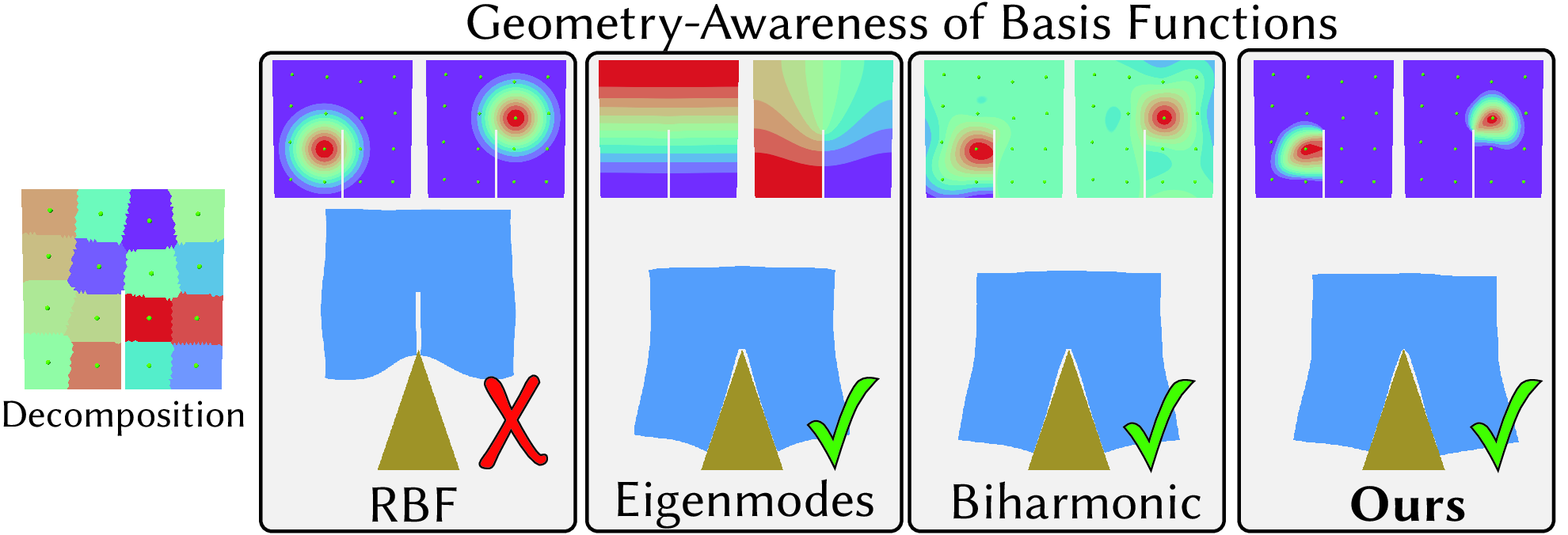}
    \caption{
        Evaluating the geometry-awareness of different basis functions.
        We drop a square with a thin cut in the middle onto a spike, simulated within our framework with four different basis functions: isotropic radial basis functions, Skinning Eigenmodes \citep{benchekroun2023fast}, biharmonic coordinates \citep{wang2015linearsubspacedesign}, and our bases.
        Each method is provided the same number of degrees of freedom, and for RBF, biharmonic, and ours we use the same handle positions.
        Radial basis functions have no awareness of the underlying geometry, leading to basis support that spans over the gap, and as a result it is unable to stretch around the spike.
        Skinning Eigenmodes, biharmonic coordinates, and our method all produce geometry-aware bases, and so produce the correct general deformation.
        \label{fig:geometry-aware}}
\end{figure}

\begin{figure}
    \includegraphics[width=\linewidth,keepaspectratio]{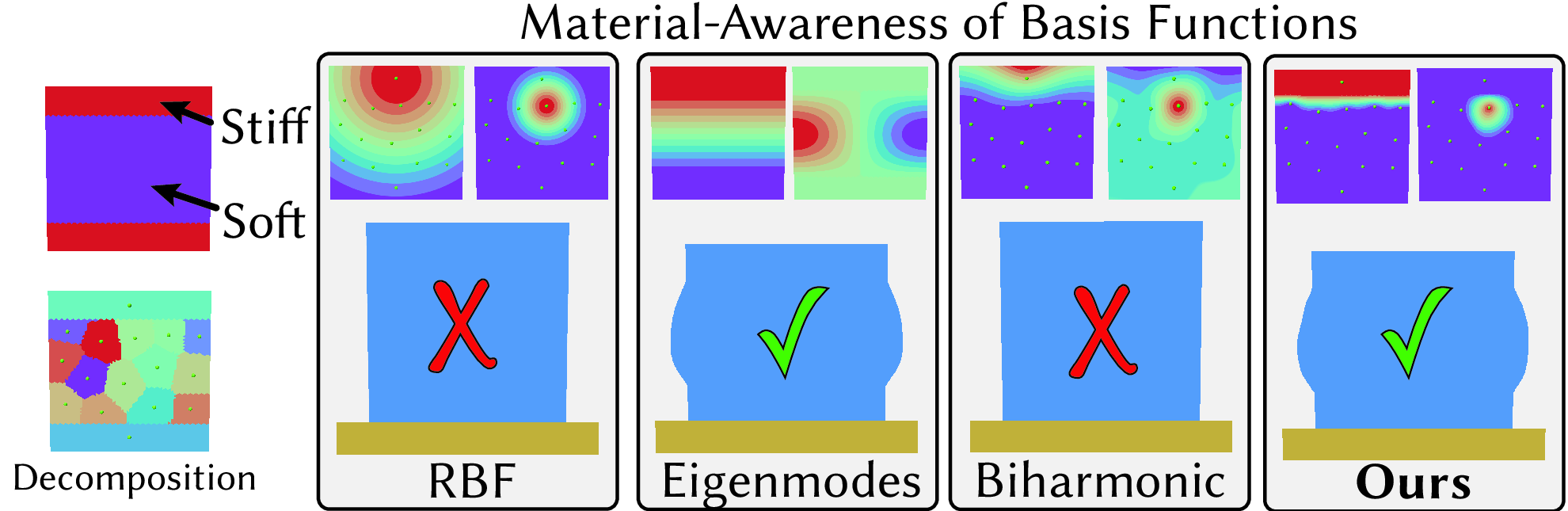}
    \caption{
        Evaluating the material-awareness of different basis functions.
        We drop a heterogeneous material square onto the ground. The square is composed of a stiff (E=1e11 Pa) top and bottom, and a soft middle (E=2e5 Pa). We simulate the square with four different basis functions: isotropic radial basis functions, Skinning Eigenmodes \citep{benchekroun2023fast}, biharmonic coordinates \citep{wang2015linearsubspacedesign}, and our bases, each with the same number of degrees of freedom.
        Radial basis functions have no awareness of the underlying material distribution, leading to basis support that spans over the soft and stiff region, and as a result the simulation is "locked" in that the soft region cannot deform.
        Biharmonic coordinates does not consider material parameters in its construction, and so similarly it produces locked deformation as a result of not conforming to material boundaries.
        Skinning Eigenmodes and our method both produce material-aware bases, and so produce the correct general deformation with bulging in the soft region.
     \label{fig:material-aware}}
\end{figure}

\begin{figure}
    \includegraphics[width=\linewidth,keepaspectratio]{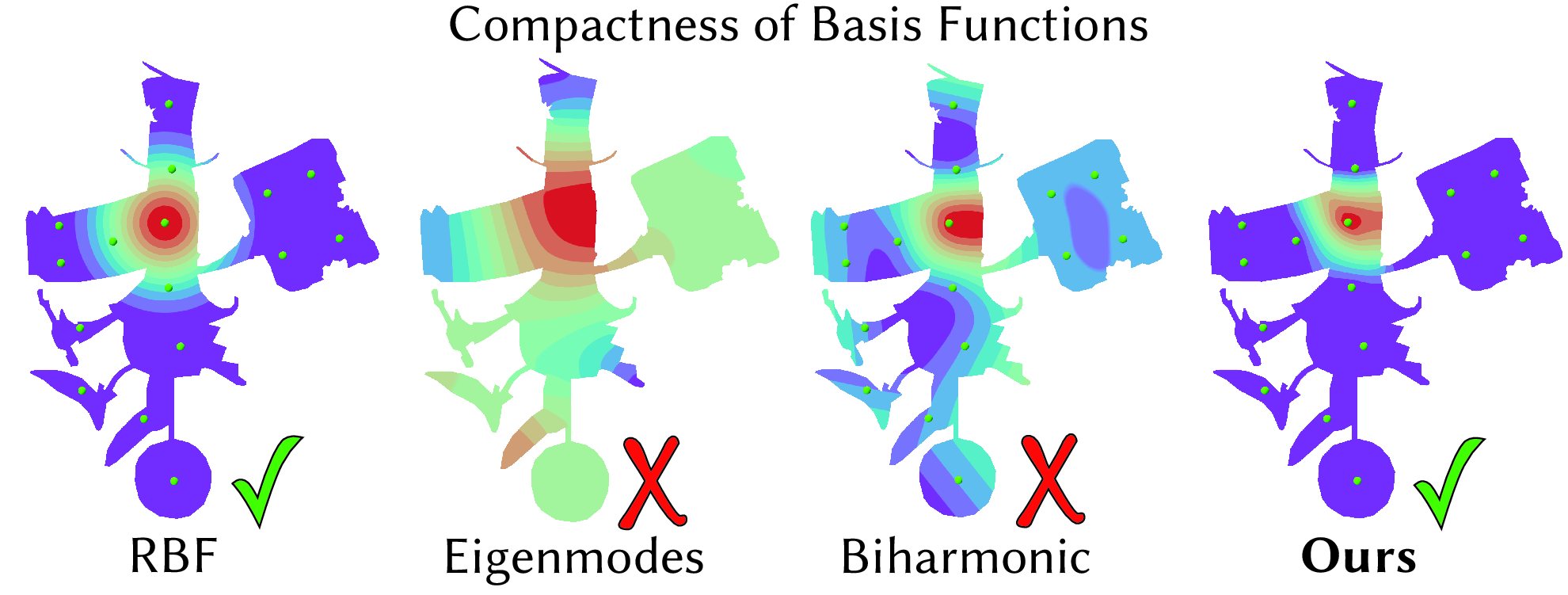}
    \caption{
        Commonly uses basis functions lack compact support, a necessary property for efficient numerical methods. Here we show the support of RBFs, Skinning Eigenmodes \citep{benchekroun2023fast}, biharmonic coordinates \citep{wang2015linearsubspacedesign}, and our method. We see only RBF and our method have compact support. Eigenmodes are naturally global, and biharmonic coordinates decay from the source point, but do not have compact support.
     \label{fig:compact-support}}
\end{figure}

\paragraph{Meshless Methods}
Handle-based methods can also be considered as a class of  meshless methods. Generally these methods are used as the sole discretization within a simulation, such as in smoothed particle hydrodynamics, where radial basis functions (RBFs) \citep{koschier2022sphsurvey} are used to discretize the domain. These RBFs are generally simple isotropic kernels with no awareness of the underlying geometry and materials, leading to artifacts when used as a coarse subspace for full-space dynamics (see figure \reffig{material-aware} and \reffig{geometry-aware}). However, with compact support (\reffig{compact-support}) these methods are well-suited for efficient parallel algorithms. Recent work improving RBF with high-order precision \citep{chen2017rkpm} improves basis function reproducibility, but does not address the interpolation weights that lead to the aforementioned artifacts. In graphics \citet{martin2010elastons} use high-order DOF on RBF bases, similar to handle-based methods, but still experience the same consequences of using isotropic kernels.

\paragraph{Generalized Basis Functions} 
In graphics, methods in physics-based animation and character control have proposed interpolatation schemes that address the problems that stem from using simple radial bases.

Many of these are based on solutions to constrained PDEs, for instance \citet{joshi2007harmonic} solves Laplace equations with dirichlet constraints to construct cage-based interpolation weights for shape deformation. Extensions of this include the smoother biharmonic coordinate variants, such as \citet{jacobson2011boundedbiharmonicweights} that enforces positivity for intuitive interpolation weights in animation, and \citet{weber2012biharmonic} and \citet{wang2015linearsubspacedesign} which focus on producing linearly precise interpolation weights. These methods all produce weights that are geometry-aware (\reffig{geometry-aware}), but lack material-awareness~(\reffig{material-aware}). Additionally, these methods are typically used in very coarse discretizations, specific to animation tasks, and lack compact support (\reffig{compact-support}) necessary for efficient numerical methods and for avoiding locking during contact. \citet{rustamov2011multiscale} extends biharmonic kernels to introduce sparsity, but requires the user to tune a kernel fall-off parameter.

With the desire for local control, many alternate interpolation schemes have been proposed in the context of character and face animation. \citet{baran2007automatic} uses heat-based basis functions constructed on the surface, using an input skeleton for control. In a physics-animation setting, \citet{kim2011multidomain} computes local modes in a decomposition, but requires explicit rotation coupling between the domains for coherence. For face animation \citet{tena2011interactiveFaces} also uses region based subspaces, leading to interface artifacts \citep{neumann2013sparse}. \citet{neumann2013sparse} focuses on facial animation as well, producing face localized modes with sparsity inducing regularizer. Collectively none of these approaches offers the properties of parameter free, geometry-awareness, material-awareness and compact support, that are crucial for our approach.

\paragraph{Modal Subspaces}
An alternative approach to producing coarse discretizations is modal analysis. \textit{Eigenmodes} \citep{pentlandwilliams1989goodvibrations} are often constructed from the eigenvectors of the rest-state elastic Hessian, resulting in deformation modes that respect physical materials (\reffig{material-aware}) and the mesh structure (\reffig{geometry-aware}). Skinning Eigenmodes \citep{benchekroun2023fast,benchekroun2023subspaceMFEM} are the most applicable variant for our purposes since the algorithm directly produces eigenfunctions-based handle weights.  However, the global nature of the bases~(\reffig{compact-support}) limits the number of modes that can be deployed and causes contact locking. \citet{brandt2017compressed} extends the Eigenmodes with sparsity-inducing L1 regularization, but this requires careful tuning of the stiffness parameter (or a more expensive iterative optimization procedure). And while this approach does favor sparse modes, it cannot guarantee the locality required to avoid contact locking between disparate portions of an object.

\subsection{Fast Solvers for Elastodynamics}

\begin{figure}
    \includegraphics[width=\linewidth,keepaspectratio]{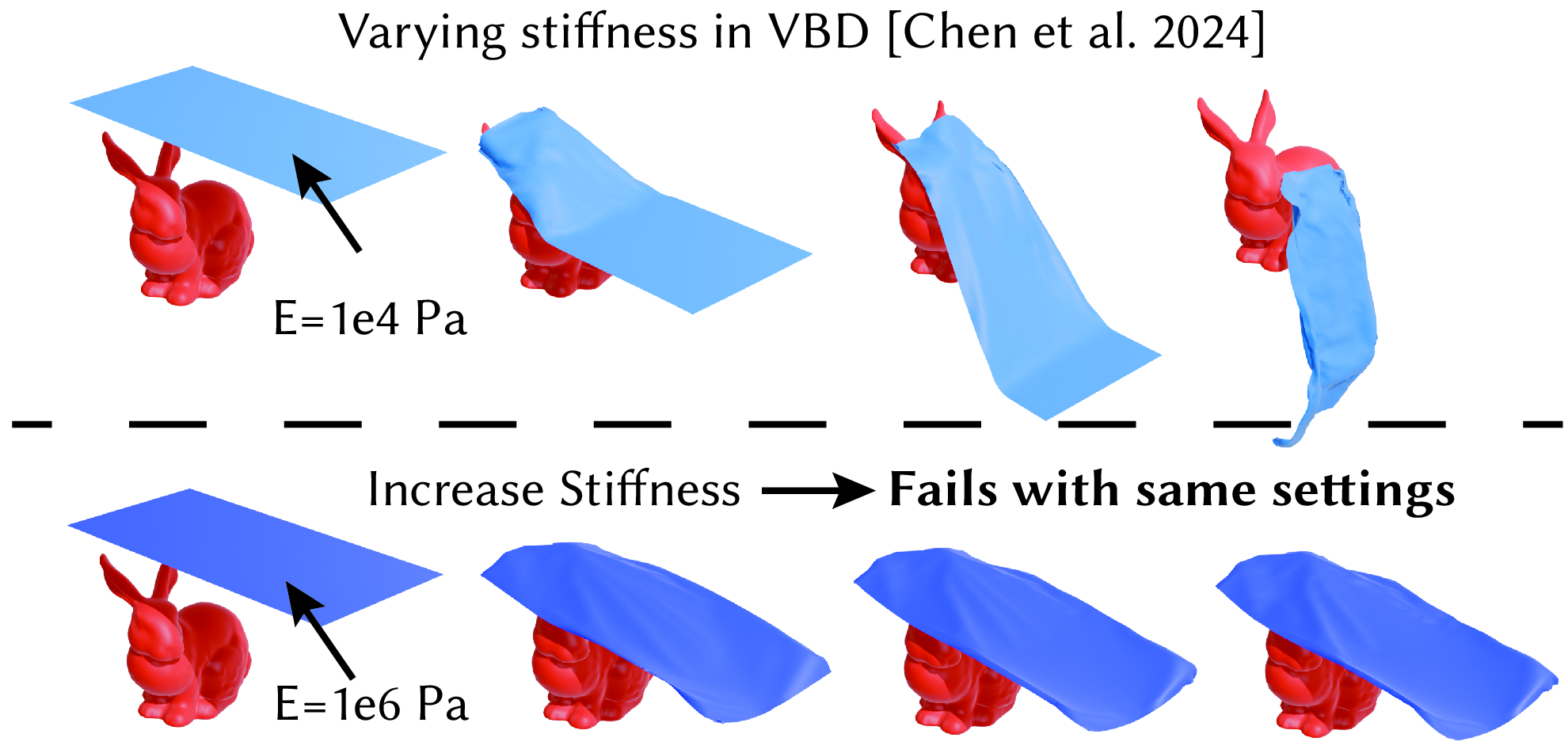}
    \caption{Coordinate Descent-type methods are highly sensitive to changes in simulation parameters. Here we simulate a cloth sheet dropping on a fixed bunny with Vertex Block Descent (VBD) \citep{chen2024vertex}. With soft materials and a low-resolution sheet, the algorithm outputs a reasonable animation. However, with unchanged solver parameters, we modify the cloth stiffness (E = 1e4Pa $\rightarrow$ 1e6Pa), and observe that the simulation produces highly damped outputs for the same number of timesteps and number of solver iterations per timestep. The performance of methods like VBD is highly parameter dependent, with the number of iterations to converge generally being a function of material stiffness and mesh resolution.
     \label{fig:vbd}}
\end{figure}

One can also accelerate simulations through improvements in the optimization loop at the heart of an implicit time integrator. Such approaches have achieved significant speedups in a number of applications. However, their performance is generally only achieved on a small set of physics regimes (typically homogenous objects with soft materials).
Like us, \citet{tiantianMultigrid2019} turn to a hierarchical iterative solver structure (multigrid in their case) for improved performance, but the lack of material-aware restriction and prolongation limits their method to homogenous objects, although they could potentially employ the basis functions we construct to alleviate this.
\citet{wu2022additiveschwarz} propose an additive Schwarz preconditioner for the conjugate gradient method which is fast, but shows convergence and peformance degradation for stiff or heterogeneous objects.
Projective Dynamics (PD)~\citep{Bouaziz2014-lo} formulates elastodynamics in terms of constraint projections, similarly to \citet{XPBD}. However both these methods heavily relax the physical model, with PD only capable of using quadratic material models. ADMM-PD \citep{overby2017admmpd} extends PD to allow for general material models, but these methods all suffer from the same issues stemming from the local-global nature of their solvers, with slowed or complete non-convergence as simulation parameters change \citep{Li:2019:DOT}. Recently, coordinate descent approaches \citep{chen2024vertex,lan2023stencil} have been proposed. However, they likewise suffer from similar resolution and stiffness dependent convergence challenges\ \citep{trusty2024tradingspaces} (see Figure \reffig{vbd}). In the next section we detail how our new basis-solver combination overcomes many of the limitations of these previous approaches.
%
%




\section{Method}

\subsection{Background}
We simulate large-deformation elastodynamics with frictional contact on simplicial meshes $\mathcal{T}$ (tetrahedra in 3D, triangles in 2D).  Applying piecewise-linear discretization we store discrete fields for position and velocity in vectors $x, v \in \mathbb{R}^{d n}$ at the $n$ vertices of the mesh in $d$-dimensional ($d$ respectively 3 or 2) space. Each time step update solve can then be cast in optimization form as
\begin{align}
x^{t+1}=\argmin_x E(x),
\end{align}
with the incremental potential (IP)\ \citep{kane2000variational,li2020IPC},
\begin{align}
\begin{split}
\label{eq:IP}
E(x) =  \> &K(x) +  \alpha h^2 \big( \Psi(x) + B(x) + D(x) \big),\\
&K(x) = \frac{1}{2} \| x - \tilde{x}^t \|_{M}^2,
\end{split}
\end{align}
formed by the weighted sum of deformation $(\Psi)$, contact barrier $(B)$, and friction $(D)$ potential energies, and an inertial weighting energy $(K)$. Choice of predictor position, $\tilde{x}^t$ (an explicit function of prior position and velocity), scaling term $\alpha \in \mathbb{R}^{+}$, and explicit update equation for velocity from optimal solution $x^{t+1}$, then jointly define the specific choice of numerical time integration method. As concrete examples, we use here implicit Euler with
\begin{align}
\tilde{x}^t=x^t+h v^t, \> v^{t+1}=\frac{1}{h}\left(x^{t+1}-x^t\right), \ \text{and} \  \alpha=1,
\end{align} 
and BDF2 with 
\begin{align}  
\tilde{x}^t &= \frac{1}{3}\left(4 x^t-x^{t-1}\right)+\frac{2 h}{9}\left(4 v^t-v^{t-1}\right), \> \alpha =4 / 9,  \\ 
a^{t+1} &=\frac{4 h^2}{9}\left(x^{t+1}-\tilde{x}^t\right), \> v^{t+1}=\frac{1}{3}\left(4 v^t-v^{t-1}\right)+\frac{2 h}{3} a^{t+1}.
\end{align}

Per timestep solves are then applied by Newton-type methods with each iteration applying a linear solve,
\begin{align}
\label{eq:full-space-linear-solve}
H d = -g,
\end{align}
where $H$ is the Hessian, $\nabla_x^2 E(x)$, of the energy, or an approximation thereof, and $g = \nabla_x E(x)$ is the energy gradient. For stiff systems, including IPC-based\ \cite{IPC} methods, this linear solve is generally a major bottleneck, often requiring expensive linear solvers (e.g., Cholesky-based) to generate quality descent directions, $d$, with ill-conditioned Hessians.

\subsection{Algorithm}
\label{sec:algorithm}

We construct a new nonlinear solver method for the efficient and high-quality approximation of timestep solutions of the IP minimization problem, \refeq{IP}. Our goals are fidelity, efficiency, reliability, and general applicability.

For efficiency, we want a \emph{fully parallelizable} method which, in turn, requires the ability to apply \emph{iterative} linear solvers. 
For reliability, we seek a method that ensures \emph{robust}, and so non-intersecting, inversion-free, stable simulation output for all timesteps.
For fidelity we require solutions that directly consider the above IP energy model, and can deliver solutions close in quality, both in terms of visual expressiveness and quantitatively, see Section\ \ref{sec:results}, to a direct minimization of the IP energy, at a fraction of the cost. 
Finally, for general applicability, we expect these properties to hold in our solver across 
widely varying simulation conditions, including extreme material stiffnesses, heterogeneities (both material and geometric), practical timestep sizes, high speeds, challenging collisions, and large applied boundary conditions.

To do so we propose a conceptually simple, three-level solver algorithm. At each level we solve Newton-type iterations with an improving approximation of the simulation domain to better capture the underlying full-resolution IP problem. 

Our first, and coarsest approximation, is simply the application of per-object affine bases\ \citep{lan2022affine}, that efficiently capture global motions, typically requiring just a single iteration to converge its stage. See Section\ \ref{sec:sparse-subspace}  for details.

Our second, and most critical level applies a custom-constructed mesh-partitioned sparse subspace basis that efficiently captures mid-scale, material-aware deformations. Critically, our subspace construction provides high-quality descent directions for the IP, with just the application of block-Jacobi preconditioned conjugate gradient (PCG). In the following sections we next focus on the construction of our basis, its corresponding mesh partitioning, and its evaluation with the IP energy. 

Finally, our midlevel subspace solves serve as a bridge to a final, efficient, full-space refinement. Here, we again apply PCG, but now with a fixed number of carefully constructed Hessian updates for a low-cost final polishing of each timestep's solution with highest-frequency enrichment. See Section\ \ref{sec:full-space-refinement} for details.

\begin{algorithm}
    \caption{Inexact Newton Solve}
    \label{alg:newton-solve}
    \KwIn{$x^t$: positions; $U_a$, $U_s$: Affine and sparse bases}
    \KwOut{$x^{t+1}$: Updated positions}
    \vspace{0.2em}
    $x \gets x^t$ \\

    \While{not converged}{
        \tcc{\textbf{Gradient and Hessian Evaluation}}
        Compute full-space gradient $g = \nabla_x E(x)$ \\
        Compute cubature Hessian $H$ (Sec.~\ref{sec:subspace-integration}) \\
        \If{\texttt{should\_update\_Hlag()}}{
            Recompute $H_{\text{lag}}$ (Sec.~\ref{sec:full-space-refinement}) \\
        }

        \tcc{\textbf{Compute Descent Direction} (see Alg.~\ref{alg:linear-solve})}
        $d_s \gets$ \texttt{SolveSubspaceDirection}$(H, g, U_a, U_s)$ \\
        $d_f \gets$ \texttt{SolveFullspaceCorrection}$(H_{\text{lag}}, g - H d_s)$ \\
        $d \gets d_s + d_f$ \\ [0.2em]

        \tcc{\textbf{Line Search and Update}}
        Perform backtracking line search to find step size $\alpha$ \\
        $x \gets x + \alpha d$ \\ [0.2em]

        \tcc{\textbf{Check termination}}
        \If{$\|d_s\| / h\numverts < \epsilon$}{
            \textbf{converged} $\gets$ \texttt{true}
        }
    }
    \Return{$x$}
\end{algorithm}

\subsection{Subspace Criteria}
As covered above, we want our mid-level sparse subspace solve to capture mid-scale system deformations in order to serve as a bridge to a final, full-space refinement that enriches with fine-scale detail. 
To maximize the effectiveness of this level, we seek a displacement basis 
that is maximally expressive for our input simulation meshes.
We argue that to fulfill this criterion, basis functions must be geometry-aware to correctly capture 
geometric and topological features of the simulation domain~(\reffig{geometry-aware}), 
be material-aware to localize deformation detail where it is needed, e.g., in relatively soft parts 
of the object~(\reffig{material-aware}), have compact support~(\reffig{compact-support}) 
to provide sparsity in system matrices and localized forces and, finally, 
be convergent, so that accuracy improves as number of DOFs increase~(\reffig{convergence}).  
To our knowledge, there is currently no basis construction method that satisfies these desiderata~(\refsec{related}).
Below we propose a two-stage algorithm 
for satisfying these criteria,
built with a mesh partitioning into material-aware subdomains (\refsec{mesh-partitioning}), 
and compactly supported biharmonic interpolation weights constructed atop that partition, introduced below.

\subsection{Subspace Construction}
\label{sec:subspace-construction}

The first two levels of our multi-level solver rely on handle-based
discretization. We define the deformed position \(\mathbf{x}=\{x,y,z\}\)
of a reference space point
\(\bar{\mathbf{x}}=\{\bar{x},\bar{y},\bar{z}\}\) using a vectorized form
of standard linear blend skinning:

\begin{equation}
	\mathbf{x} = \sum_{i=1}^n \underbrace{\phi_i\left(\bar{\mathbf{x}}\right)P_i\left(\bar{\mathbf{x}}-\bar{\mathbf{x}}	_i\right)}_{\mathcal{R}^{3\times 12}}\mathbf{q}_i,
    \label{eq:linear-blend-skinning}
\end{equation}
where for $n$ handles, \(\phi_i\left(\bar{\mathbf{x}}\right)\in\mathcal{R}\) is a scalar
weighting function, \(\bar{\mathbf{x}}_i\in\mathcal{R}^3\) is the origin
of the \(i^{th}\) handle in the reference space and
\(\mathbf{q}_i\in\mathcal{R}^{12}\) is a flattened affine matrix (the
handle itself). Each level of our mult-level solver defines a different
number of handles and computes \(\phi_i\) in a different way.
\(P_i\left(\mathbf{\bar{\mathbf{x}}}\right)\) is a monomial basis given by

\[P_i(\bar{\mathbf{x}}) = I_3 \otimes \begin{bmatrix} 1 & \bar{x} & \bar{y} & \bar{z}\end{bmatrix},\]
where \(I_3\) is the \(3\times 3\) Identity matrix and \(\otimes\) is
the Kronecker product. 

For our solver's global level, we assign a single handle to the center-of-mass of
a each object and set \(\phi_i\left(\bar{\mathbf{x}}\right)=1\) (except when there are Dirichlet boundary conditions, see \refsec{dirichlet-bc}) which gives
this handle control over global affine motions of the objects. To construct the per-object affine subspace's basis matrix, $U_a$, we stack the $3\times 12$ blocks associated with each vertex~\refeq{linear-blend-skinning} in the base mesh.

\subsection{Sparse Material-Aware Subspace}
\label{sec:sparse-subspace}
Our sparse material-aware subspace (SMS) follows a similar construction as the affine one, but with the key difference being in the weighting function $\phi_i(\vc{x})$. Among other properties, our function is designed to be compactly supported in the neighborhood of each handle. To do this we require, as input, a mesh partitioning to define the support domain of the basis functions. We cover this below in \refsec{mesh-partitioning}.
First, we seek a basis function that effectively interpolates between handles, and is enforced to be zero outside of its support region. Our partitioning segments the mesh into $m$ partitions, and, on each partition, we build a basis function centered on its centroid $\bar{\vc{x}}_i$.

\begin{figure}
    \includegraphics[width=\linewidth,keepaspectratio]{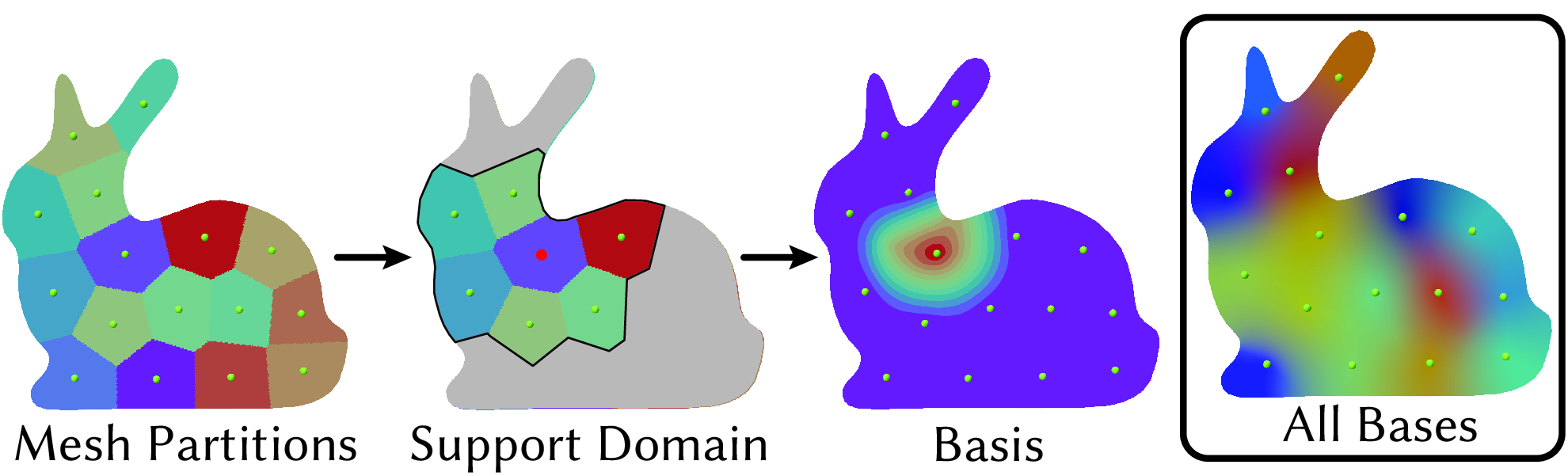}
    \caption{
        The end-to-end process of constructing the SMS bases. First, we partition the mesh, placing handles at the centroids of each partition. Next, for each partition we collect the local neighborhood around the partition to form a subdomain. And on this subdomain, we solve for our bases. The far right image shows all the bases blended together (each with a different color), showing the smooth transition between the bases.
     \label{fig:basis-construction}}
\end{figure}

For each partition, we form a subdomain consisting of the partition and its immediate neighbors (see \reffig{basis-construction}). Neighbors are those that share a face with the partition. This collection of partitions is denoted $\mathcal{T}_i$. To define our scalar weighting function, $\phi_i(\bar{\vc{x}})$, we use a biharmonic solve, similar to that in \citet{wang2015linearsubspacedesign}, but extended to guarantee compact support and account for material properties.
Rather than using the standard biharmonic operator, we use a stiffness-weighted biharmonic operator so that the resulting bases are sensitive to the material distribution.

Our biharmonic solve is
\begin{equation}
    K u = 0 \> \text{in } \mathcal{T}_i,
\end{equation}
where $K$ is the stiffness-weighted bilaplacian operator discretized in $\mathcal{T}_i$. We subject our solution, $u$, to the Dirichlet constraints
\begin{enumerate}
    \item $u = 1$ at the centroid of the partition, $\bar{\vc{x}}_i$,
    \item $u = 0$ at the centroids of all neighboring partitions in $\mathcal{T}_i$,
    \item $u = 0$ on $\partial \mathcal{T}_i \setminus \partial \mathcal{T}$, i.e., the internal boundary of the subdomain excluding the boundary of the mesh.
\end{enumerate}
These boundary conditions ensured that the biharmonic solve is compactly supported within $\mathcal{T}_i$, vanishing at its boundaries, and blends smoothly with neighboring subdomains. The third set of constraints excludes the boundary of the full domains, as we want to guarantee non-zero weights at the mesh boundaries.

\begin{figure}
    \includegraphics[width=\linewidth,keepaspectratio]{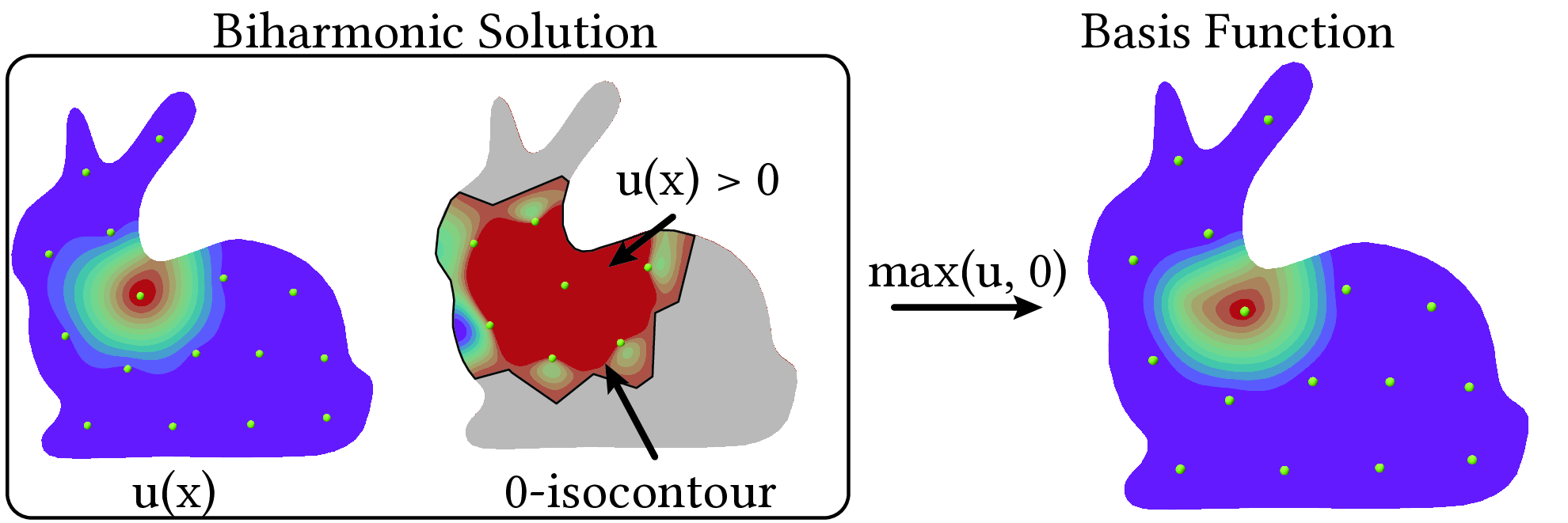}
    \caption{
        Our bases are derived from the solution to a biharmonic equation, and in order to define the support of the basis functions, we clip the solution to retain only its non-negative values. The left image shows the full biharmonic solution, and the middle shows the $0$-isocontour as well as the negative region (red is positive, and blue is negative). This $0$-isocontour serves as the support boundary for the basis functions. The right image shows the clipped solution with weights decaying from $1$ at the centroid to $0$ at the support boundary.
     \label{fig:support_boundary}}
\end{figure}

Recall that our goal is to construct a basis that decays smoothly from $1$ at the cluster centroid to $0$ at some notion of support boundary. Ideally this boundary would be with respect to neighboring centroids. The biharmonic solution naturally defines a potential boundary, but extends past neighboring centroids to an "in-between" region, between them and $\partial \mathcal{T}_i$, with negative values. We thus clip the biharmonic solution to just its non-negative values, $u_\text{pos} = \max(0, u)$, and so obtain a smooth support that transitions cleanly near a boundary between the neighboring centroids (see \reffig{support_boundary}).

\begin{figure}
    \includegraphics[width=\linewidth,keepaspectratio]{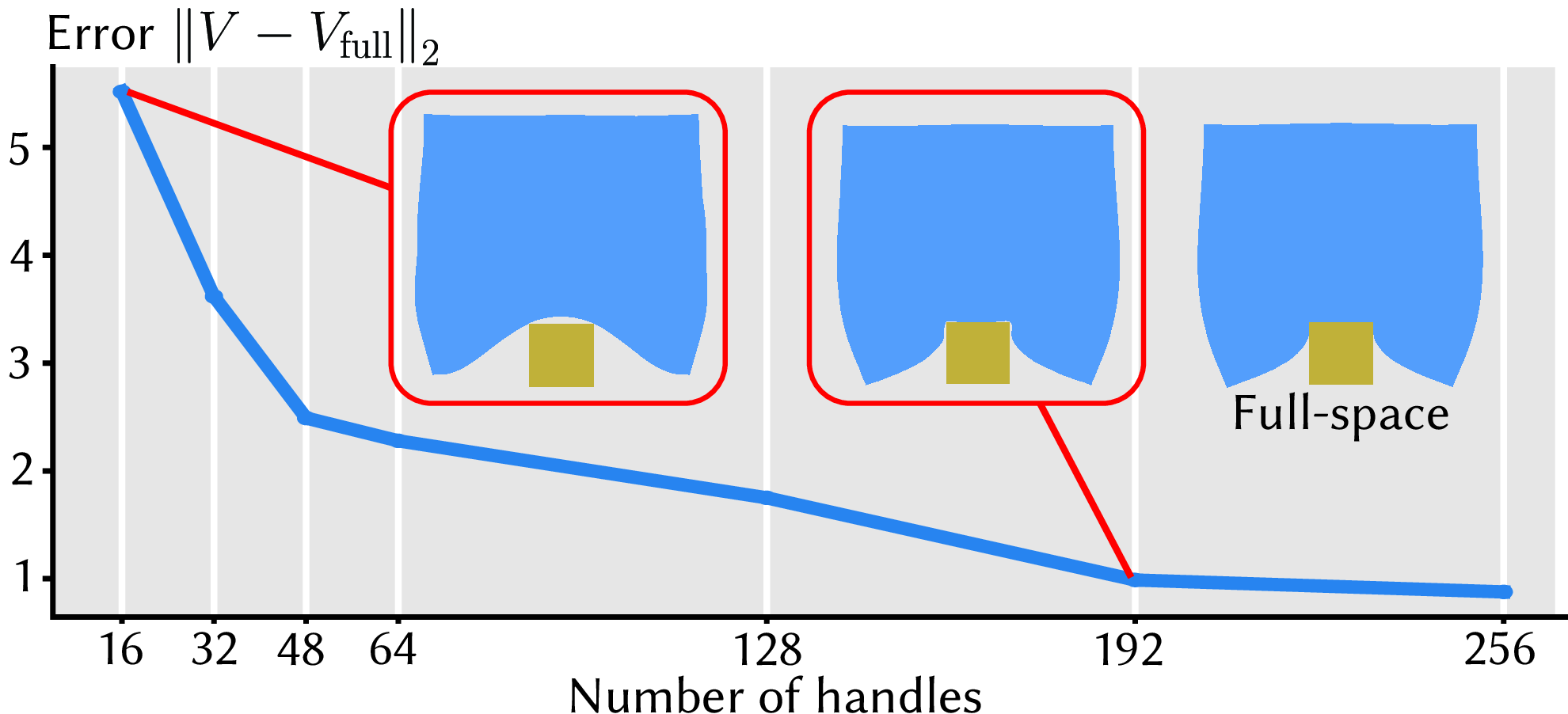}
    \caption{
        With partition of unity, the SMS bases converge to the full-space solution as the number of handles increases.
        Here we simulate a soft square impacting a fixed square (yellow) in a single high-speed timestep, and measure the error in the displacement with respect to the full-space solution.
        As we increase the number of handles, the error decreases and qualitatively the simulation at 192 handles is nearly indistinguishable from the full-space solution.
     \label{fig:convergence}}
\end{figure}

While this clipped solution, $u_\text{pos}$, is at least $C0$ smooth and compactly supported, our disjoint, per-partition solves sacrifice linear precision offered by global biharmonic coordinates.
To address this, we enforce a partition of unity (POU) across the overlapping support regions. This is done by normalizing the clipped weights at each vertex,
\begin{equation}
    \phi_i(\bar{\vc{x}}) = \frac{u_{\text{pos},i}(\bar{\vc{x}})}{\sum_{j=1}^m u_{\text{pos},j}(\bar{\vc{x}})},
\end{equation}
where $u_{\text{pos},i}(\bar{\vc{x}})$ is the clipped biharmonic solution for the $i$-th partition. With POU, our final weights are guaranteed to reproduce constant functions, and because we use affine degrees of freedom, we recover linear precision. This property is important to guarantee that the subspace basis is convergent under refinement, meaning we approach the full-space solution as we add more handles (see \reffig{convergence}).

Constructing our basis functions from these weights follows the same process as the affine subspace. For each vertex covered by the support of a handle, we assign a $3 \times 12$ block in the basis:
\begin{equation}
    U_s(\bar{\vc{x}}) = \begin{bmatrix}
        \phi_1(\bar{\vc{x}}) P_1(\bar{\vc{x}}) & \phi_2(\bar{\vc{x}}) P_2(\bar{\vc{x}}) & \cdots & \phi_m(\bar{\vc{x}}) P_m(\bar{\vc{x}})
    \end{bmatrix},
\end{equation}
where this $3 \times 12$ block is typically sparse due to the compact support of the basis functions. Conceptually, these blocks can be stacked together to form the full sparse subspace basis, $U_s \in \R^{dn \times r_s}$, where $r_s$ is the number of sparse DOF (e.g. $12 \times \text{num handles}$ in 3D). In practice (see below) we never need to assemble this matrix.

\subsubsection{Dirichlet Boundary Conditions}
\label{sec:dirichlet-bc}
For simulations with prescribed Dirichlet boundary conditions (DBCs), the basis functions (both the affine and SMS) must be updated to be made compatible with the DBCs.
We achieve this by modifying the bases' weighting functions.

\begin{figure}
    \includegraphics[width=\linewidth,keepaspectratio]{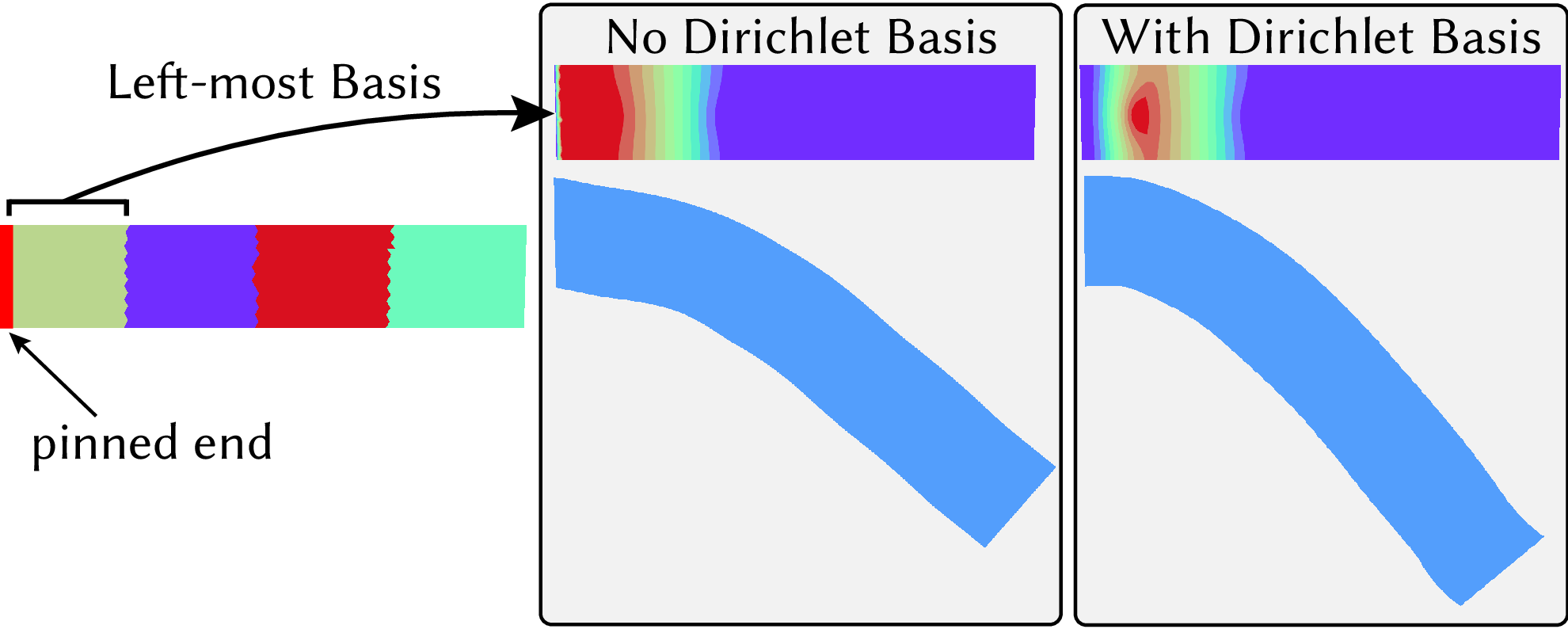}
    \caption{
        The SMS bases are modified to be compatible with dirichlet boundary conditions (DBC) by introducing an additional basis for the DBC points.
        Here we have a cantilevered beam with 4 handles. Without DBC compatibility, the SMS bases are discontinuous at the pinned end, preventing the beam from smoothly bending. By extending the SMS bases to account for the DBC during the biharmonic solve and the POU normalization, we get smooth deformation at the boundary and greater total deflection. 
       \label{fig:dbc_beam}}
\end{figure}

First, for the SMS subspace, we modify the biharmonic solve, enforcing additional Dirichlet constraints $u=0$ at all DBC vertices. This ensures that the basis functions vanish at these points. Next, we need to account for the DBCs in our POU normalization. To do this we introduce a basis function corresponding to the DBCs, $u_\text{DBC}(\bar{\vc{x}})$. This basis is built equivalently as the bases for the handles, except the biharmonic solve enforces $u=1$ at DBC vertices, and $u=0$ at neighboring centroids (corresponding to partitions that contain any DBC vertices).
This ensures smooth support at the DBC vertices, and that the POU normalization is consistent with the DBCs as well:
\begin{equation}
    \phi_i(\bar{\vc{x}}) = \frac{u_{\text{pos},i}(\bar{\vc{x}})}{\sum_{j=1}^m u_{\text{pos},j}(\bar{\bar{\vc{x}}}) + u_\text{DBC}(\bar{\vc{x}})}.
\end{equation}

\subsection{Mesh Partitioning}
\label{sec:mesh-partitioning}
\danny{Ty: we have multiple versions here but its not clear when/how/where each variant is used - this needs to be clarified. See questions below and missing value needed too.}
With our above SMS subspace bases defined we next cover here the construction of the mesh partitioning which defines their support. 
Like the SMS basis functions themselves, our partitioning should \textit{also} be both geometry- and material-aware. These properties are essential to ensure that basis functions are localized to regions expected to exhibit similar deformations.

Beyond these requirements, we additionally favor partitions that are approximately equal in size (as scaled by stiffness) and as convex as possible.
This encourages regularity in the partitioning, avoiding highly irregular basis function domains which, in turn, can lead to imbalanced computational loads (i.e. if partitions may have much more neighbors than others).

There is a wealth of methods for mesh partitioning, most methods do not satisfy these properties. For instance, standard methods like METIS\ \citep{karypis1997metis} will yield geometry-aware partitions of roughly equal volume, but have no notion of material properties. On the other hand, methods like spectral clustering can yield material-aware partitions, but tend to produce non-connected partitions. \ty{A short didactic figure would be good here I think.}


\begin{figure}
    \includegraphics[width=\linewidth,keepaspectratio]{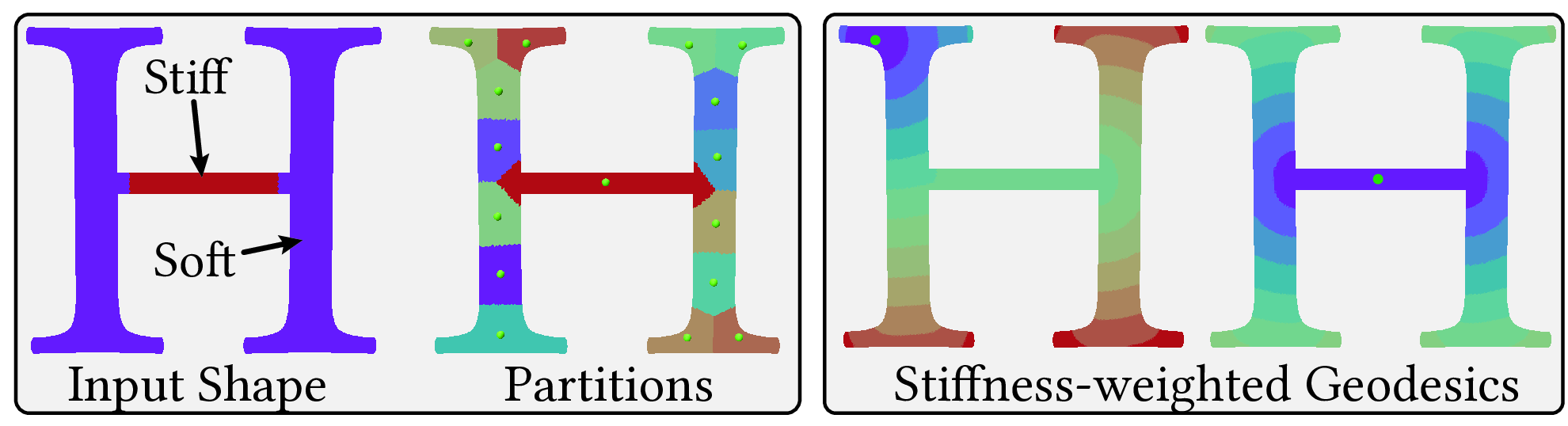}
    \caption{  
    We construct material-and geometry-aware mesh partitions. On left we show an `H' input geometry with thin features and a material distribution with a stiff inner region and soft elsewhere. The result partitions roughly conform to this distribution and respect the geometry.
    This is achieved through clustering based on stiffness-weighted geodesics (right). With these stiffness-geodesics, distances in the middle of the 'H' in the stiff region move faster, and as a result more area is assigned to clusters in this region.
     \label{fig:clustering}}
\end{figure}

\subsubsection{Stiffness-and Geometry-Aware K-means}
\label{sec:stiffness-weighted-heat-geodesics}
To build a decomposition via an appropriate set of clusters, we perform K-means clustering by Lloyd relaxation, applying an extension of the heat diffusion-based geodesics from \citet{crane2013geodesics}.
To make these distances material-aware we utilize a stiffness-weighted Laplacian, corresponding to diffusion with a spatially varying diffusion coefficient.
With heat flowing "faster" in stiff regions, the K-means clustering will naturally form larger clusters in these regions, and smaller clusters in softer regions.
This can be seen as a more general variant of the compliance distances used in \citet{faure2011sparsemeshless}.

Following standard Lloyd relaxation, we require an input number of clusters, $k$. We initialize these points via kmeans++\ \citep{arthur2007kmeans}, using the same above stiffness-weighted geodesic distance metric.

\subsubsection{Accounting for Large Material Change}
\label{sec:partitioning-jump-penalty}
Applying just stiffness weighting can still lead to decompositions that will not closely adhere to material boundaries, and this, in turn, can lead to a dearth of clusters assigned to softer regions as they are compressed out. To address this issue, we modify the Laplacian for our heat-distance calculation to include a jump penalty. Wherever a node touches elements with differing materials, we reassign the stiffness on these faces to a small value (we choose 0.1 to 0.5 $\times$ the softest material). This effectively slows down the heat in these regions, encouraging tighter clusters. \danny{We always use this or just sometimes?} \ty{we always use this. Values go in range 0.1-0.25 in our examples (0.1 for hoop, should be 0.25 for others)}

\subsubsection{Accounting for Local Minima}
\label{sec:partitioning-merging}
On complex geometry, K-means clustering can frequently converge to poor quality, local minima with highly imbalanced clusters. To remedy this, we further modify K-means. Rather than using a fixed, $k$, number of clusters, we densely sample the shape. Then, during the iterative procedure (every 5-10 \danny{What's the choice for this to?}\ty{10 iterations on all examples, 5 on kooshball} iterations), we remove clusters that are “too small”. We measure this condition by the ratio between a cluster’s volume and the median cluster volume. If this measure is too small, we mark the cluster for potential deletion. When such clusters are found we then delete the $n$ smallest clusters that satisfy this criterion.  \danny{what is $n$?} \danny{Again: always use this or just sometimes?} \ty{always use this. $n$ is 5 on all examples except koosh ball, where i use $n=30$ (so i can terminate solver faster) } \danny{Ok - gonna need a table for the different param choices used in this section per example at end of paper.}

\subsection{Per-Iteration Linear Solves}
\label{sec:iterative-linear-solve}

At each iteration of multi-level timestep solver we proceed in two stages: first we compute a subspace descent direction, and then, second, we apply a full-space refinement to enrich this  direction. The overall per-iteration solve is outlined in Alg.~\ref{alg:linear-solve}.

\begin{algorithm}
    \caption{Linear Solves: Subspace and Full-Space}
    \label{alg:linear-solve}
    \SetKwFunction{SolveSubspace}{SolveSubspaceDirection}
    \SetKwFunction{SolveFullspace}{SolveFullspaceCorrection}
    \SetKwProg{Fn}{Function}{:}{}

    \Fn{\SolveSubspace{$H$, $g$, $U_a$, $U_s$}}{
        \tcc{\textbf{Affine subspace solve}}
        Solve $(U_a^\top H U_a) q_a = U_a^\top g$ using PCG \\
        $d_s \gets U_a q_a$ \\
        $r \gets g - H d_s$ \\ [0.2em]

        \tcc{\textbf{Sparse subspace correction}}
        Solve $(U_s^\top H U_s) q_s = U_s^\top r$ using PCG \\
        $d_s \gets d_s + U_s q_s$ \\
        \Return{$d_s$}
    }
    \vspace{0.5em}
    \Fn{\SolveFullspace{$H_{\text{lag}}$, $r$}}{
        Solve $H_{\text{lag}} d_f = r$ using PCG \\
        \Return{$d_f$}
    }
\end{algorithm}

\subsubsection{Subspace Solve}
\label{sec:subspace-solve}
Our subspace direction solve is split into a first-level affine subspace solve followed by our mid-level SMS subspace solve. Both solves use the same Hessian approximation, $H$ (see \refsec{subspace-integration}), and both are solved using PCG, each solved to a relative residual of $10^{-4}$ throughout.

Directly assembling the subspace sandwich matrices, $U_a^T H U_a$ and $U_s^T H U_s$, necessary for our subspace solves is generally prohibitively expensive on account of 
support of these bases, so we apply a partially matrix-free approach. For a basis $U$, we compute the  $(U^T H U) q$ matrix-vector product as a sequence of three products:
\begin{align*}
    w &= Uq, \quad \; \text{(lift to full-space)}, \\
    y &= H w, \quad \text{(application of Hessian)}, \\
    z &= U^T y \quad \text{(restriction to subspace)}.
\end{align*}
This avoids the need for explicit matrix-matrix multiplication, and is readily parallelizable as it only requires a parallel matrix-vector product against the Hessian and the basis matrix.\dave{we never build U right?} \ty{no, we bsaically have a custom sparse matrix rep though, i have something akin to a CSR-like structure and have hand-rolled kernels for the sparse-matrix vector product against the basis.} \danny{Worth adding ;)}
For the subspace solves, block sizes for the Jacobi preconditioner  correspond to the number of DOFs per handle (e.g., in 3D $12 \times 12$ for affine blocks). 

\subsubsection{Full-space Refinement}
\label{sec:full-space-refinement}

\begin{figure}
    \includegraphics[width=\linewidth,keepaspectratio]{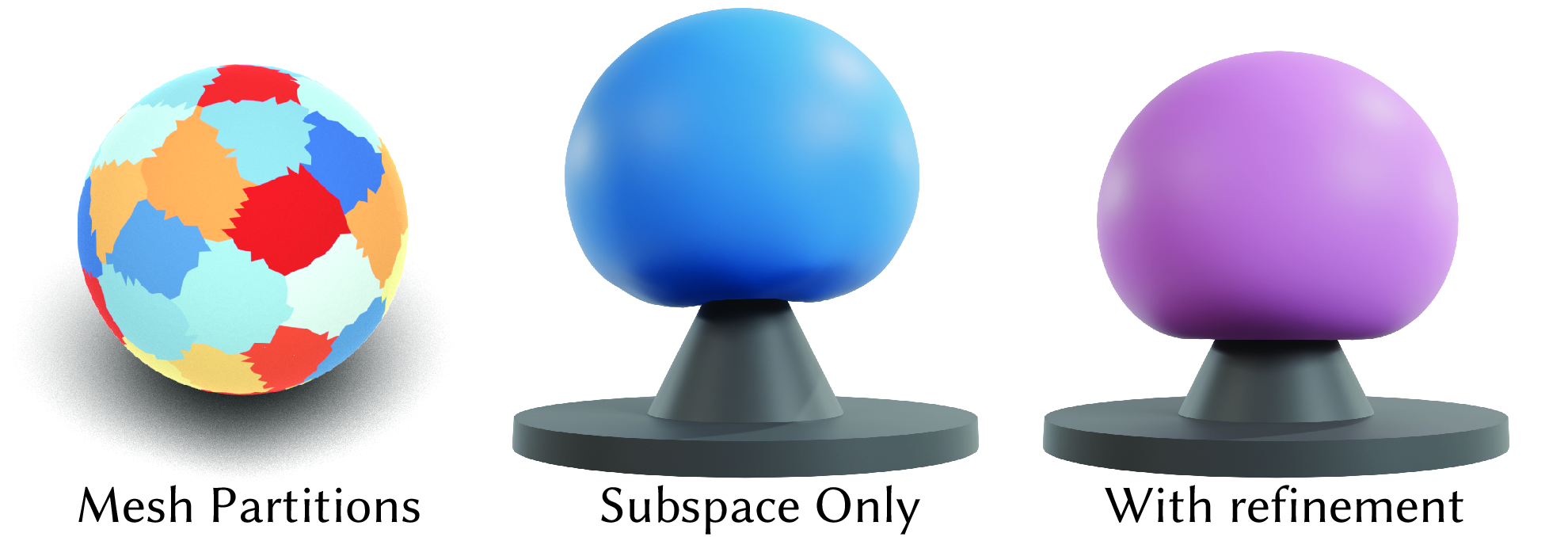}
    \caption{
    Full-space refinement is essential recovering high-frequency deformation, such as the sharp deformation due to this sphere's impact with a spike. We simulate our method with and without full-space refinement on this sphere with 64 SMS bases.
    Without refinement (middle) the sphere compresses much less than the the version with refinement (right) due to the inability of the pure subspace solution to resolve high-frequency details.
     \label{fig:refinement}}
\end{figure}

As discussed in \refsec{algorithm}, our full-space refinement only targets resolution of high-frequency details not captured by the subspaces. With our goal of producing fast approximations, we want to avoid as much expensive full-space computation as possible. So, for this solve, we perform just a fixed precomputed (see below for our calculation of iteration count per system) number of CG iterations with
\begin{equation}
    H_{\text{lag}} d_f = r,
\end{equation}
where $H_{\text{lag}}$ is an approximate, lagged Hessian (see below). Here the right-hand side, $r$, is the residual from the subspace solve, $r = g - H d_s$, for the subspace direction, $d_s$. Our lagged Hessian, $H_{\text{lag}}$, is sparse, so parallelization is straightforward for the CG solve, only requiring a parallel sparse matrix-vector product with it. 

\paragraph{Preconditioning}
For our refinement solve we use diagonal (Jacobi) preconditioning. Full-space Jacobi preconditioners generally struggle with propagating information across the mesh. However, the prior bases themselves already account for multiscale effects. In a sense our subspace solves roughly serve as preconditioners to capture low-to-medium frequency details, so that the full-space solve focuses primarily on enriching with local details. 

\paragraph{Hessian Lagging}
As we take a small number of full-space iterations, the bottleneck for full-space refinement is Hessian assembly, particularly of our elastic-energy terms, which require visiting all volume elements. To alleviate this cost, we apply a lagged approximation of the elastic energy Hessian. Importantly, we always freshly recompute all other energy Hessians, e.g., our contact barrier terms. This guarantees that our Hessian approximation is appropriately scaled in the presence of new contact stencils.

Lagged Hessians can easily become stale, degrading the quality of descent directions \citep{brown2013lowrank}, which leads to small line-search step sizes. We always update the elastic Hessian at the beginning of a timestep, but this can still easily become stale within a timestep solve involving large deformation.
To determine when to trigger a Hessian update, we use the step size from backtracking line search as a proxy for Hessian quality. Small step sizes may indicate inaccurate curvature estimates, but they can also result from higher-order effects or emerging contacts. Therefore, we conservatively track a moving average of recent step sizes and trigger a Hessian update only when this average drops below a threshold. Further, to avoid excessive updates, we then recompute the elastic Hessian at most once every five iterations. 

\paragraph{Choosing the Number of CG Iterations}
We use a fixed number of CG iterations (20–40) for full-space refinement across all examples. This choice is motivated by the need to resolve fine-scale features that are not captured by the SMS bases.
Since Jacobi-preconditioned CG propagates information at a rate tied to the mesh resolution, a reasonable estimate for the required number of iterations is the number of edge hops needed to traverse the diameter of a subspace partition.
We measure this by computing the maximum number of edge hops from each partition's centroid to its farthest vertex, average this quantity over all partitions. Empirically we find that this value is always close to the range of 20-40. In practice we then simply bin this simulation parameter by choosing polishing iterations as the closer value of either 20 or 40 to this measure.

\subsection{Subspace Hessian Integration}
\label{sec:subspace-integration}

Our subspace linear solves described in the last section apply sparse matrix-vector products with the subspace basis and the full-space Hessian.
While the cost of multiplying by the basis depends on the number of handles and basis sparsity, the Hessian-vector product scales with the number of elements in the full mesh. As such, using the full Hessian is not practical, and yields a subspace product that is effectively equivalent to the cost of a full-space product.

To decouple subspace solve cost from mesh resolution, we approximate Hessian integration using \textit{cubature}.
Cubature selects a subset of elements with positions $\{\bar{\vc{x}}_i\}_{i=1}^n$ and weights $\{w_i\}_{i=1}^n$ to approximate integrals over the domain
\begin{equation}
    \int_{\Omega} f(\bar{\vc{x}}) \, d\Omega \approx \sum_{i=1}^n w_i f(\bar{\vc{x}}_i).
\end{equation}
This reduces the cost of fully integrating functions over all mesh elements, requiring function evaluations at only a smaller subset of elements.

For our subspace linear solves, we use cubature specifically to approximate just the elastic energy Hessian contributions, which is the most costly term in the Hessian-vector product. It typically contains the most number of nonzeros by a large margin in its fully integrated form.

\paragraph{Moment Fitting}
We compute cubature weights using \textit{moment fitting}, which selects weights $\{w_i\}$ such that the above approximation is exact for a target function space, typically polynomials such as:
\begin{equation}
    \{1, \bar{x}, \bar{y}, \bar{z}, \bar{x}^2, \bar{x}\bar{y}, \bar{y}^2, \dots\},
\end{equation}
or orthogonal bases like Chebyshev and Legendre polynomials.

While this is effective for smooth, polynomial functions, classical moment fitting has no awareness of the actual basis functions used in our subspace methods, which are non-polynomial. As a result, it becomes difficult to choose an appropriate polynomial degree: low-order polynomials result in under-integration (low quality subspace directions), and high-order polynomials lead to dense integration (high cost).

Standard moment fitting fails to adapt to basis complexity and how it varies through the mesh. For example, in our SMS subspaces the basis functions depend on material parameters: stiff regions tend to produce nearly constant bases, while soft regions lead to more complex, varying ones.
An ideal cubature scheme should reflect this variation: allocating fewer integration points where the basis is simple, and more where it is complex. Classical moment fitting offers no such adaptivity.

\subsubsection{Modal Moment Fitting}

To address this, we extend moment fitting to incorporate the subspace basis functions themselves as the target function space.
To motivate this extension, consider a point-wise energy we wish to integrate over the reference domain, given by $E(x\left(\bar{\vc{x}}\right))$. The deformed position $\vc{x}$ can be represented as $\vc{x} = \bar{\vc{x}} + U\left(\bar{\vc{x}}\right)\vc{q}$, where $U\left(\bar{\vc{x}}\right)$ is a vector of $m$ subspace values at $\bar{\vc{x}}$. 
Substituting, and applying Taylor expansion in terms of $\Delta\vc{x} = U\left(\bar{\vc{x}}\right)\vc{q}$ yields:
\begin{equation}
    E(\bar{\vc{x}} + U\vc{q}) = E(\bar{\vc{x}}) + \vc{q}^T U^T \nabla_x E + \frac{1}{2} \vc{q}^T U^T \nabla_{x}^2 E U\vc{q} + \mathcal{O}(\|\vc{q}\|^3).
\end{equation} 
This involves linear and quadratic combinations of basis functions, and so to accurately compute integrated energy, gradient, and Hessian values, we must accurately integrate these combinations.

Let $U_s = [u_1, \dots, u_{r_s}]$ denote the SMS subspace basis.
For a degree-$p$ approximation, we aim to integrate all monomials formed from basis functions up to degree $p$. For $p=2$ this includes
\begin{equation}
    \int_{\Omega} 1 \, d\Omega, \quad
    \int_{\Omega} u_i(\bar{\vc{x}}) \, d\Omega, \quad
    \int_{\Omega} u_i(\bar{\vc{x}}) u_j(\bar{\vc{x}}) \, d\Omega, \quad \forall i, j.
\end{equation}
More generally, let $\mathcal{I}_p$ be the set of multi-indices representing
all polynomial combinations of the basis functions up to degree $p$.
Then our moment fitting conditions become:
\begin{equation}
    \int_{\Omega} \prod_{i \in \alpha} u_i(\bar{\vc{x}}) \, d\Omega = \sum_{k=1}^n w_k \prod_{i \in \alpha} u_i(\bar{\vc{x}}_k), \quad \forall \alpha \in \mathcal{I}_p.
\end{equation}

\begin{figure}
    \includegraphics[width=\linewidth,keepaspectratio]{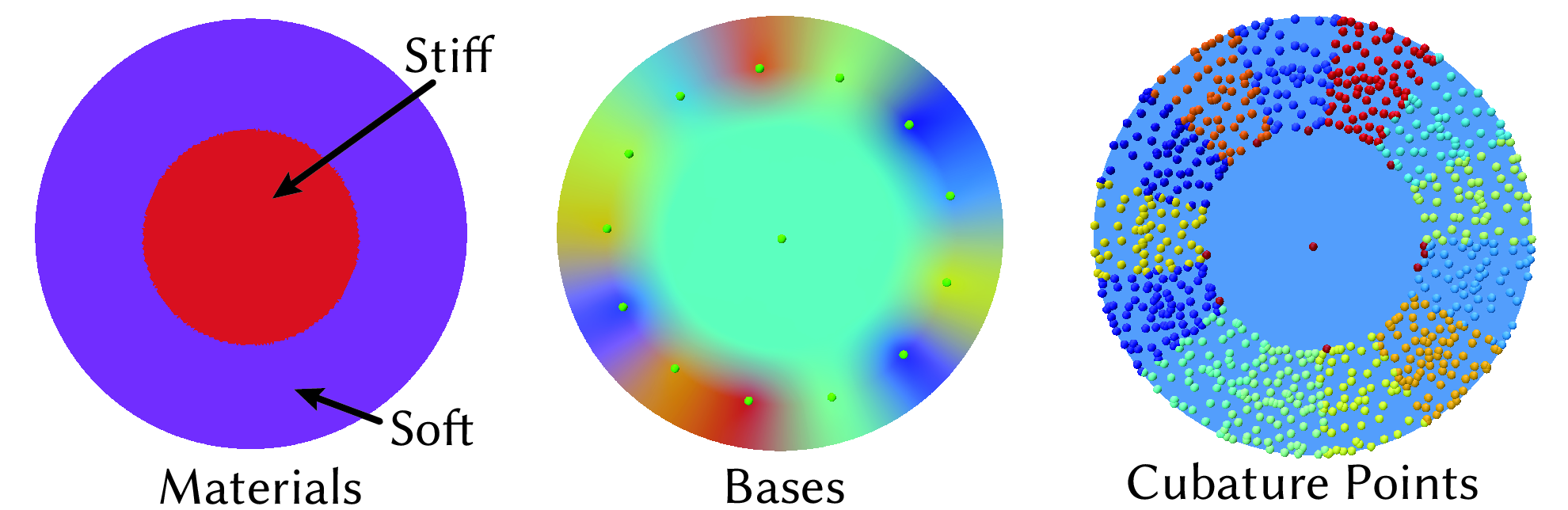}
    \caption{
        Our modal moment fitting cubature weights adapt to the complexity of subspace basis functions. Stiff regions yield simpler basis functions and require fewer cubature points; soft regions exhibit more variation and require denser sampling.
     \label{fig:cubature}}
\end{figure}

This \textit{modal moment fitting} adapts automatically to local basis complexity.
In stiff regions, basis functions are nearly constant, yielding redundant constraints that can be satisfied with fewer cubature points.
In soft or complex regions, more variation in the basis necessitates denser sampling (see \reffig{cubature}).

\subsubsection{Moment Fitting Solver}
To apply this fitting scheme, we use our mesh partitioning structure from \refsec{mesh-partitioning} to build integration complexes.
Within each partition, we gather the subset of basis functions that intersect this partition (those with nonzero support). We restrict these functions to this partition, and solve for a set of cubature points and weights that satisfy the moment fitting conditions on this subset of basis functions.

To compute the cubature weights for each partition, we solve a linear system,
\begin{equation}
    A w = b,
\end{equation}
where $A \in \mathbb{R}^{m \times n}$ is the matrix of basis function evaluations (for $n$ samples), $w \in \mathbb{R}^n$ is the vector of cubature weights, and $b \in \mathbb{R}^m$ is the target integral values for each polynomial combination in $\mathcal{I}_p$.
Since the basis functions $u_i$ are defined at mesh vertices, we approximate their values at the barycenters by averaging the basis values at each element's vertices. This gives us a per-element basis function evaluation.

The target integrals $b$ are computed by integrating these basis functions combinations over the partition domain $\Omega$:
\begin{equation}
    b_{\alpha} = \int_{\Omega} \prod_{i \in \alpha} u_i(X) \, d\Omega.
\end{equation}
We discretize this integral by evaluating the basis functions on all
partition elements (at the barycenters), and summing the values weighted by
the element volumes. 

%
%
We solve this linear system using a non-negative least squares (NNLS) solver. Non-negativity is essential for the positive definiteness of the resulting cubature-based Hessian. Additionally, NNLS yields sparse solutions \citep{bro1997nnls} with the number of nonzero entries in $w$ roughly equal to the number of constraints ($m = |\mathcal{I}_p|$). The actual sparsity will depend on the rank of the fitting matrix, $A$, which depends on the complexity of the bases (polynomial combinations of a single constant basis will yield a rank-1 matrix and so 1 sample). This sparsity is crucial to ensure an efficient integration scheme that adapts to the basis complexity.

\begin{algorithm}
    \SetAlgoLined
    \SetKwInOut{KwIn}{Input}
    \SetKwInOut{KwOut}{Output}
    \KwIn{Subdomain $\Omega$, Subspace $U$}
    \KwOut{Cubature weights $w$}
    Sample $n$ cubature points from $\Omega$ \\
    Computed the integral constraints $b$\\
    Evaluate basis functions $U$ at the points to construct $A$\\
    Build $A$ and solve $A w = b$ using non-negative least squares\\

    \While{$\|A w - b\| / \|b\| > \epsilon$}{
        Sample $n$ additional cubature points from $\Omega$\\
        Rebuild $A$ and solve $A w = b$ using nnls\\
    }
    \Return $w$
    \caption{Modal Moment Fitting with Iterative Cubature}
    \label{alg:iterative-cubature}
\end{algorithm}

%
In practice, we construct the cubature scheme iteratively (see Alg.~\ref{alg:iterative-cubature}).
Starting with an initial set of $n_{\text{init}}$ randomly sampled elements, we iteratively add $n$ more points until the relative residual $\|A w - b\| / \|b\|$ falls below a tight tolerance (we use $10^{-9}$).
New points are sampled probabilistically, based on the current residual projected onto the basis functions of all unsampled elements.
Specifically we compute the residual $r = A w - b \in \mathbb{R}^m$.
Then for each remaining candidate element $k$, we compute its contribution, $\rho_k = \|A_{\cdot,k}^T r \|^2$, where $A_{\cdot,k}$ is the column vector of basis evaluations for element $k$. We then normalize $\rho_k$ across all candidates to form a probability distribution and sample $n$ new cubature points according to this distribution and repeat.

\subsection{Tying It All Together: Multilevel Timestep Solves}

With our mesh partitioning (\refsec{mesh-partitioning}), SMS bases (\refsec{subspace-construction}), and cubature sampling (\refsec{subspace-integration}) now constructed, we have all ingredients necessary to build out our multilevel timestep solve algorithm.
We begin simulation with our precomputed affine and SMS subspace basis matrices, $U_a$ and $U_s$, respectively.  At start of each time step solve, $t+1$, we begin with  prior \emph{full space} states $x^t, x^{t-1}, \cdots$, $v^t, v^{t-1}, \cdots$, as required by our applied time-integration method. We then build our corresponding IP energy (Equation\ \ref{eq:IP}) for minimization and start our Newton-type multlevel iterations. 

\paragraph{Search Directions} At each iteration of the timestep solve we first compute the IP's full-space gradient, $g$, cubature Hessian (\refsec{subspace-integration}), $H$, and update our lagged Hessian, $H_{lag}$, if triggered by our criteria (\refsec{full-space-refinement}). We then apply our subspace direction solves using the first two levels, affine and SMS subspace, of our solver (Algorithm \ref{alg:linear-solve}) to compute our subspace direction $d_s= U_a q_a + U_s q_s$, followed by our full-space refinement to get our fullspace correction, $d_f$ (\refsec{iterative-linear-solve}).

\paragraph{Linesearch} To better-approximate the full-space IPC solution, we then perform filtered linesearch on our full-space-corrected search direction, $d = d_s + d_f$ on our full-space objective. Following \citet{li2020IPC} we first filter for intersection with CCD (and non-inversion checks as energy-appropriate) to truncate the initial step size so that we can guarantee penetration-free (and injective) steps. We then follow this with backtracking line search on the full-space IP energy (Equation\ \ref{eq:IP}).

\paragraph{Termination} For termination, we then measure the potential improvement in our \emph{subspace} to further decrease our full-space objective. With our goal of applying our custom-constructed subspaces to primarily model dynamics, we terminate when our \emph{subspace} Newton decrement, $d_s$, satisfies $\|d_s\|/ h \numverts < \epsilon$. Here we use Li et al.'s\ \shortcite{li2020IPC} rescaling by the number of vertices and timestep size to measure termination relative to mesh resolution and timestep size (in units $m/s$).
Our temination criteria then effectively measures how much error, as measured in by the full-space gradient, can be projected into our subspace. In turn the amount of work we choose to apply in each Newton solve is based on the quality of our subspace to capture the full-space model's dynamics.

\section{Results}
\label{sec:results}

Our algorithm is implemented in C++, using Eigen~\cite{guennebaud2010eigen}
for linear algebra and is parallelized with Intel TBB~\cite{pheatt2008intel}
and OpenMP~\cite{dagum1998openmp}.
Timing results are reported on a machine with a 32-core AMD Ryzen Threadripper and 512 GB of RAM.
All simulations use non-inverting Neo-hookean material model with Backwards Euler or BDF2 time integration, and contact is modeled with IPC barrier functions.
All Newton solves are terminated when the criteria satisfies a tolerance of $1e-3$.
Scene parameters are summarized in Table~\ref{table:scene_statistics_extended}. \ty{todo} \ty{dont forget to report cubature ratio} \ty{oh oops i forgot this}
All comparisons are performed within the same framework with the same test bed.
Collision detection for full-space and subspace use the same framework, including a TBB parallelized sparse voxel grid. Descent directions are guaranteed to be interpenetration-free, using ACCD \citep{CIPC}.

In mesh partitioning, there are several user parameters in addition to the initial number of partitions. For our stiffness-aware K-means, we use a jump penalty at material boundaries. In elements at material boundaries, we reassign their stiffness to be a constant factor $0.1$ to $0.5$ $\times$ the softest material. In our heterogeneous material examples, we use $0.1$ for the basketball hoop (\reffig{basketball-hoop}), $0.25$ for stressball (\reffig{stressball}), and $0.5$ for the shoe (\reffig{shoe}). In our K-means algorithm, we also perform cluster pruning when clusters become too small, which is common for complex geometries like the kooshball. For all examples, we use an imbalance ratio of $1.75$, meaning that if the median cluster volume is $1.75 \times$ a cluster's volume, then that cluster is a candidate for pruning. We perform pruning every $10$ Kmeans iterations, for all examples (except the kooshball where we merge every $5$ iterations).

\begin{table*}[]
    \rowcolors{2}{white}{cyan!25}
    \caption{Statistics and parameters for various scenes. $|V|$, $|F|$, and $|T|$ represent the number of vertices, faces, and tetrahedra, respectively. $|U|$ is the number of subspace degrees of freedom, $E$ is the range of Young's Modulus (in Pascals), $\mu$ is the friction coefficient, $\Delta t$ is the time-step size in seconds, TI indicates the time integration scheme used (IE is implicit Euler), \# CG iters is the number of full-space refinement iterations, and \# Cub. Pts is the number of cubature points used in the subspace Hessian integration.}
    \label{table:scene_statistics_extended}
    \begin{tabular}{l|c|c|c|c|c|c|c|c|c|c|c}
    Scene             & Figure       & $|V|$ & $|F|$ & $|T|$ & \# Cub. Pts & $|U|$ & $\Delta t$ (s) & $\mu$ & $E$ (Pa)      & TI    & \# CG iters \\ \hline
    Basketball hoop   & \reffig{teaser}       & 83k   & 160k  & 229k  & 94k      & 2k    & 0.01           & 0.1   & 5e5--1e10      & BDF2  & 20 \\
    Basketball        & \reffig{pcg-cost}     & 61k   & 124k  & 168k  & 33k      & 1.5k  & 0.01           & 0.5   & 1e7--1e9       & BDF2  & 20 \\
    Kooshball         & \reffig{kooshball}    & 150k  & 245k  & 450k  & 13k      & 4.8k  & 0.01           & 0.2   & 5e5            & BDF2  & 40 \\
    Shoe              & \reffig{shoe}         & 200k  & 167k  & 875k  &    10k        & 756   & 0.01           & 0.5   & 1e6--1e8       & BDF2  & 20 \\
    Roller            & \reffig{DBC-roller}   & 110k  & 85k   & 516k  & 34k      & 2.1k  & 0.01           & 0.1   & 2e4--1e6       & IE    & 20 \\
    Stressball        & \reffig{stressball}   & 100k  & 20k   & 539k  & 99k      & 3k    & 0.001          & 0.5   & 4e5--1e8       & BDF2  & 20 \\
    Jello (hires)     & \reffig{jello-anvil}  & 148k  & 37k   & 813k  & 19k      & 1.5k  & 0.005          & 0.1   & 4e5--1e9       & BDF2  & 20
    \end{tabular}
\end{table*}

\begin{table*}[]
    \rowcolors{2}{white}{cyan!25}
    \caption{Comparison of IPC and subspace solver performance. Times are in seconds, and speedup is measured as the ratio of IPC time to subspace time after a fixed number of timesteps. We choose the timesteps with the most complex deformation and number of contacts, and compare speedup over a subset of the early steps as the dynamics will eventually diverge. For the Basketball example, statistics are computed over time steps 70–100 when rim contact begins.}
    \label{table:solver_speedup}
    \begin{tabular}{l|c|c|c|c|c}
    Scene                   & Figure       & Timesteps & IPC-LLT Time (s) & Subspace Time (s) & Speedup \\ \hline
        Basketball hoop             & \reffig{teaser}         & 30         & 3128.91      & 262.61             & 11.91   \\
    Stressball             & \reffig{stressball}     & 30         & 4329.12      & 566.43             & 7.64    \\
    Roller                 & \reffig{DBC-roller}     & 50         & 4393.73      & 328.31             & 13.38   \\
    Kooshball (40 CG)      & \reffig{kooshball}      & 30         & 18474.18     & 1882.33            & 9.81    \\
    Kooshball (10 CG)      & \reffig{kooshball}      & 30         & 18474.18     & 7180.79            & 2.57    
    \end{tabular}
\end{table*}

\begin{figure}
    \includegraphics[width=\linewidth,keepaspectratio]{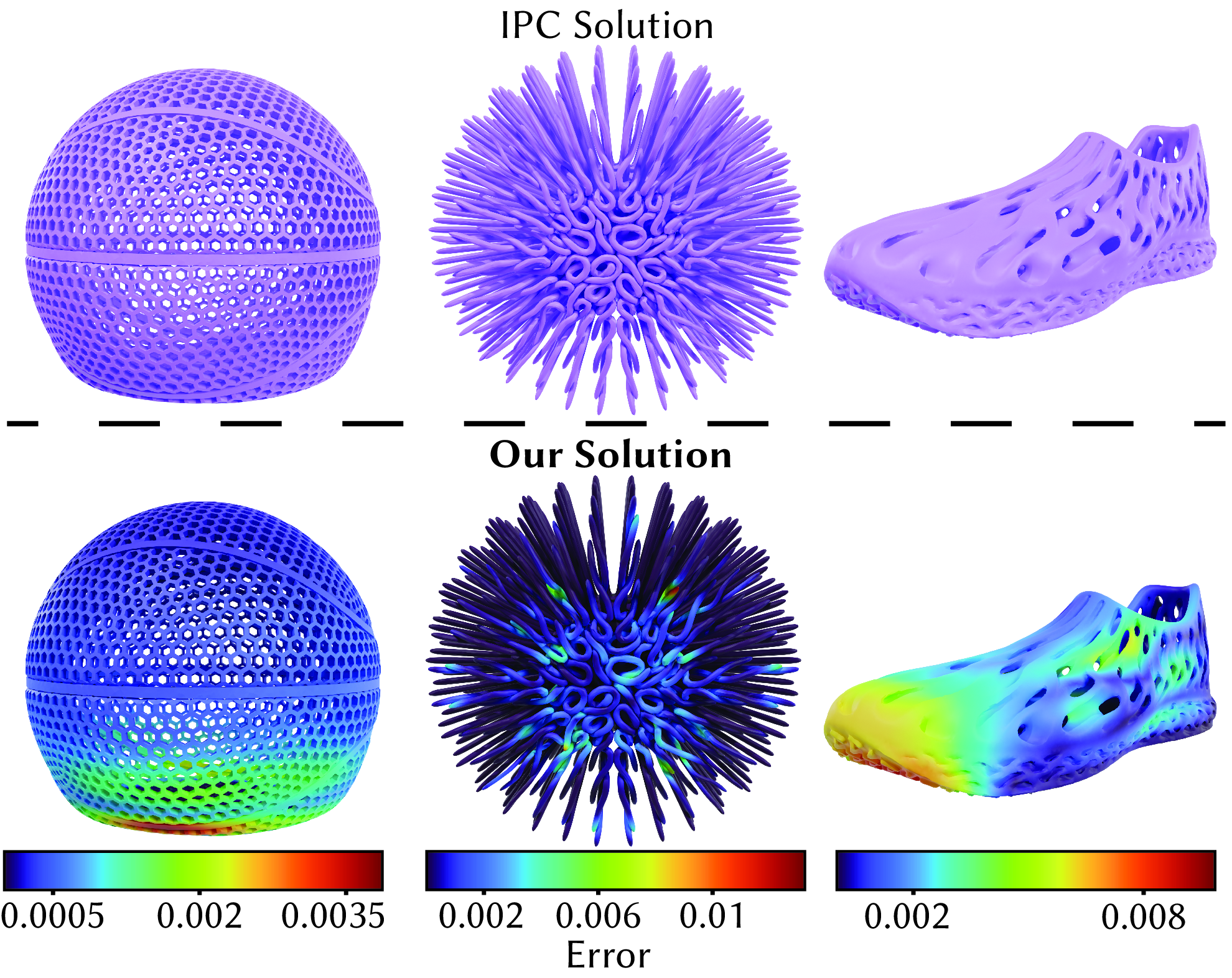}
    \caption{
        Our method often produces solutions that are visually indistingushable from the full-space IPC solution, and with measurably small error. Here we simulate the basketball, kooshball, and the heterogeneous shoe for one timestep with full IPC and our method. Both simulations start at the same state, at a high cost frame with many contacts.
        We visualize the output of ours and full-space IPC, noting the per-vertex errors on our solution.
        We measure error through the per-vertex norm of the displacement difference and divide these quantities by the bounding box diagonal of the scene, giving us a unitless relative error measure.
        As seen in the scale, the error is small for all simulations with the largest being 0.01 for the kooshball.
     \label{fig:ipc-error}}
\end{figure}

\subsection{Baseline IPC Comparisons}

\paragraph{Solution Quality} Here we simulate a lattice-design basketball, a kooshball, and a heterogeneous material shoe, choosing a large-deformation contacting timestep as \emph{same} initial conditions for bootstrapping a timestep solve with full IPC and our method. In \reffig{ipc-error} we visualize the output of our method and full-space IPC, noting the extreme visual similarities of both solutions and plotting the per-vertex errors on our solution. We measure error through the per-vertex norm of the displacement difference and divide these quantities by the bounding box diagonal of the scene, giving us a unitless relative error measure that we see is small across all simulations, with a largest error of 0.01 for the kooshball.

\begin{figure*}
    \includegraphics[width=\textwidth,keepaspectratio]{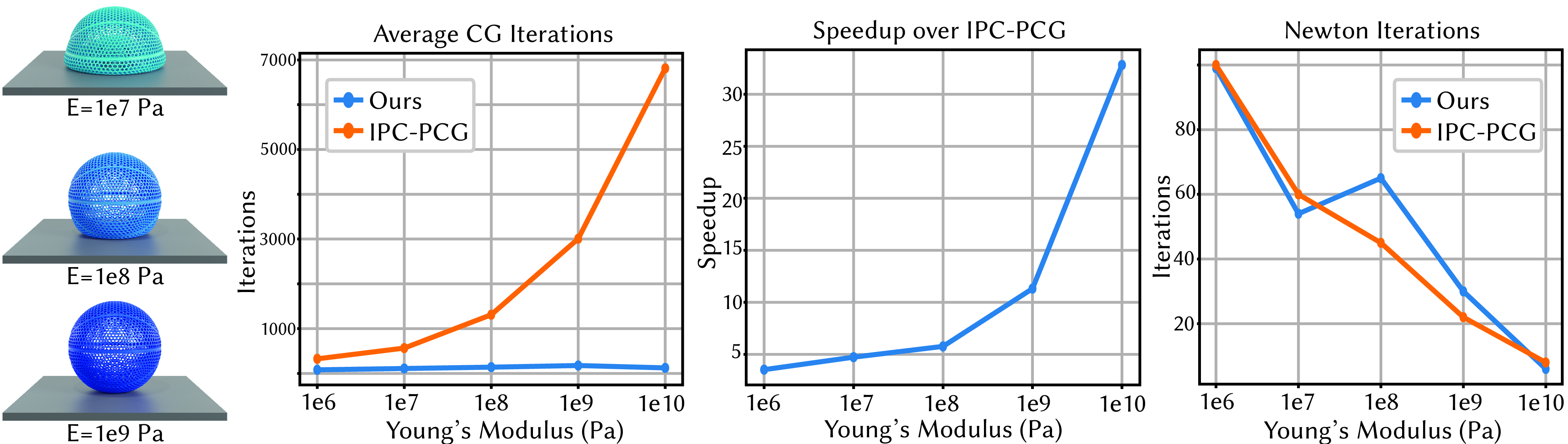}
    \caption{
        Newton-PCG is sensitive to material parameters.
        We drop the basketball from the teaser onto a plane with different materials ranging from E=1e6 to E=1e10 Pa.
        We take a frame from each simulation at the initial point of contact, and solve it with our methods and Newton-PCG (Jacobi-preconditioned CG).
        We report the average number of CG iterations over the timestep solve for both methods, the number of Newton iterations taken, and the total runtime.
        Our method reports consistent 100-200 CG iterations per step, while the Newton-PCG method's iteration count explodes as the material stiffness increases, approaching 10,000 iterations per solve at E=1e10 Pa. This leads to a timestep solve speedup over Newton-PCG of 40x.
     \label{fig:pcg-cost}}
\end{figure*}

\paragraph{Performance} We get a 7-13$\times$ speedup compared to a IPC-LLT and up to 40$\times$ speedup over IPC-PCG (see below). We summarize our method's comparative performance and timing in Table\ \ref{table:solver_speedup} and \reffig{pcg-cost}.

\subsection{Generality}
Our algorithm is general in that we offer reliable and consistent performance across a wide range of materials, material distributions, and geometries. 

\paragraph{Material Stiffness}
As a key feature our method enables the use of a simple, Jacobi-preconditioned CG for all our underlying linear solves. In turn, our subspace design and multi-level algorithm, requires fewer iterations to converge, even as materials are varied. In contrast, as we see in \reffig{pcg-cost}, standard full-space Newton PCG solver's convergence breaks down significantly and rapidly as material stiffnesses increase.

\begin{figure}
    \includegraphics[width=\linewidth,keepaspectratio]{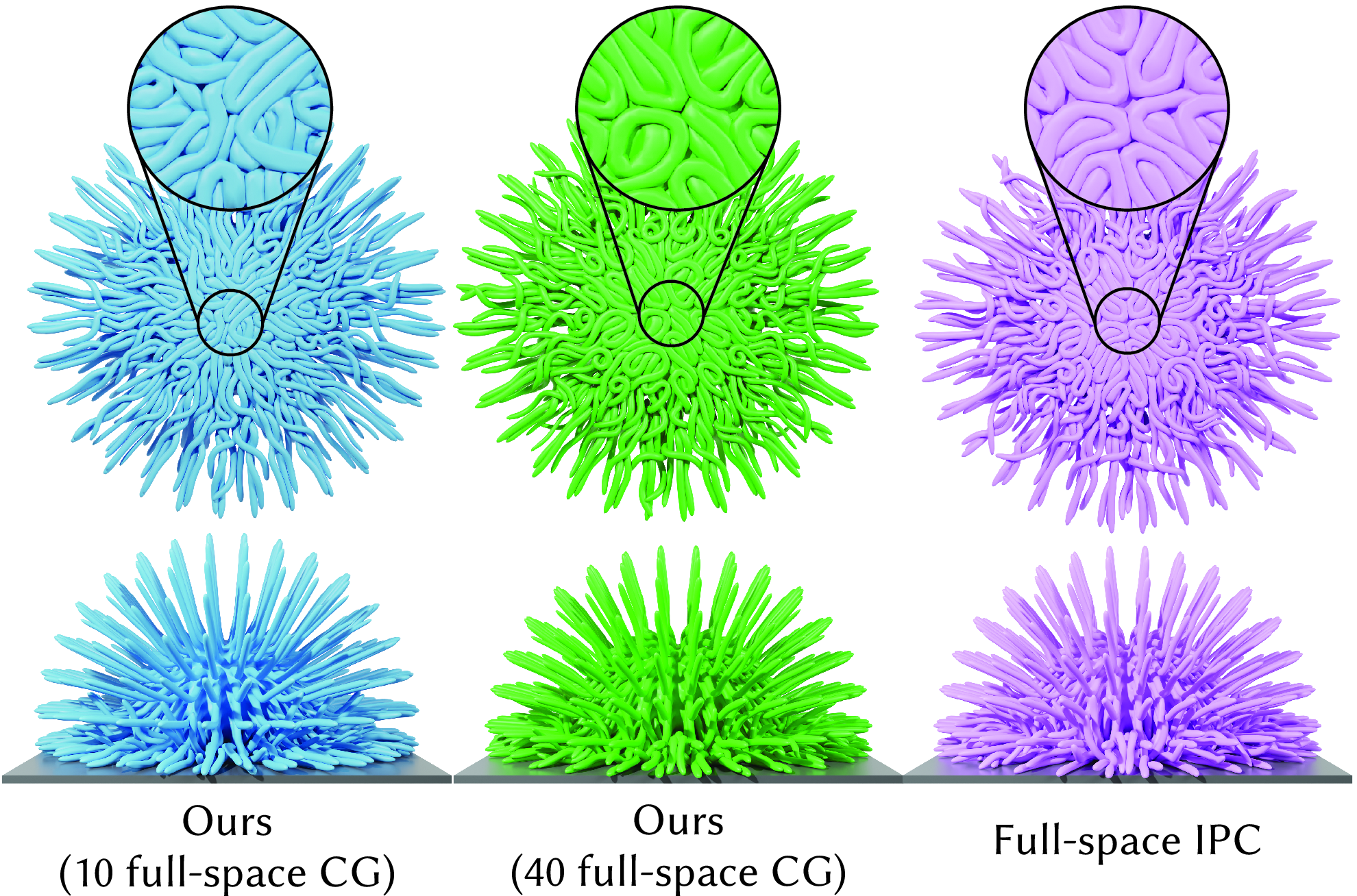}
    \caption{
        Here we drop a "kooshball" at 5 m/s onto a plane.
        We show the output of our method with 10 full-spacew refinement CG iterations (left), 40 refinement iterations (middle), and the full-space IPC solution (right).
        For this problem with complex geometry and local deformation, the effect of additional CG iterations is dramatic, and the 40 CG version is visually indistinguishable from the full-space solution (with the total runtime cost 8x faster, see video).
        10 CG version (left) noticeablly lacks the complex interlocking strands, and interestingly, the total runtime cost is much slower than the 40 CG version. This highlights how additional full-space work can accelerate the full solve, particularly when much of the dynamics is local deformation.
     \label{fig:kooshball}}
\end{figure}

\begin{figure}
    \includegraphics[width=\linewidth,keepaspectratio]{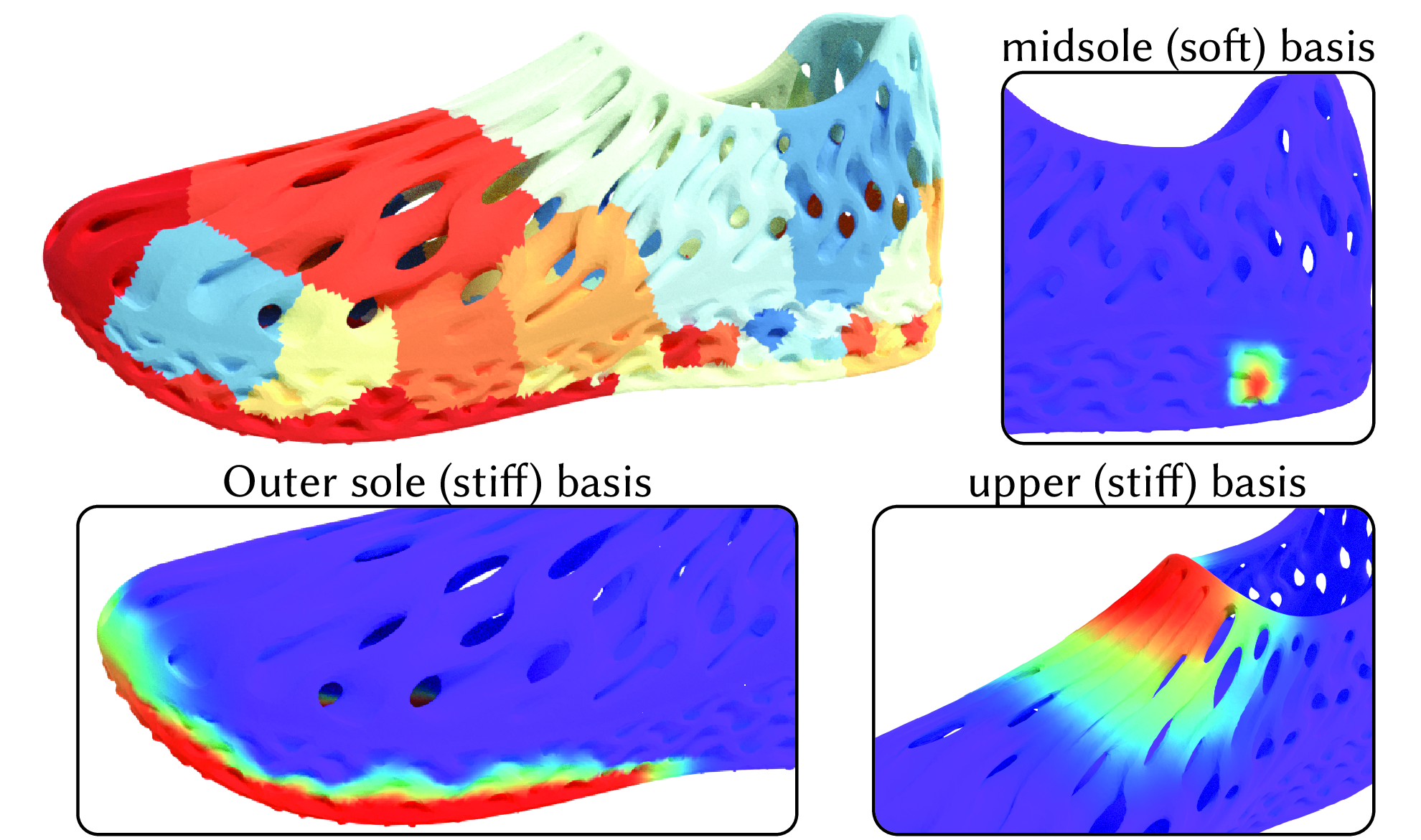}
    \caption{
        The mesh partitioning and several bases of a heterogeneous material shoe are shown. The shoe is composed of three materials: a soft midsole (E=1e6 Pa), a stiff outer sole (E=1e8 Pa), and a slightly stiff upper (E=1e7 Pa). The resulting partitioning and bases respects this material distribution, assigning large partitions (and so bases with large support) to the stiff components, and small partitions to the soft midsole.
     \label{fig:shoe_basis}}
\end{figure}

\begin{figure}
    \includegraphics[width=\linewidth,keepaspectratio]{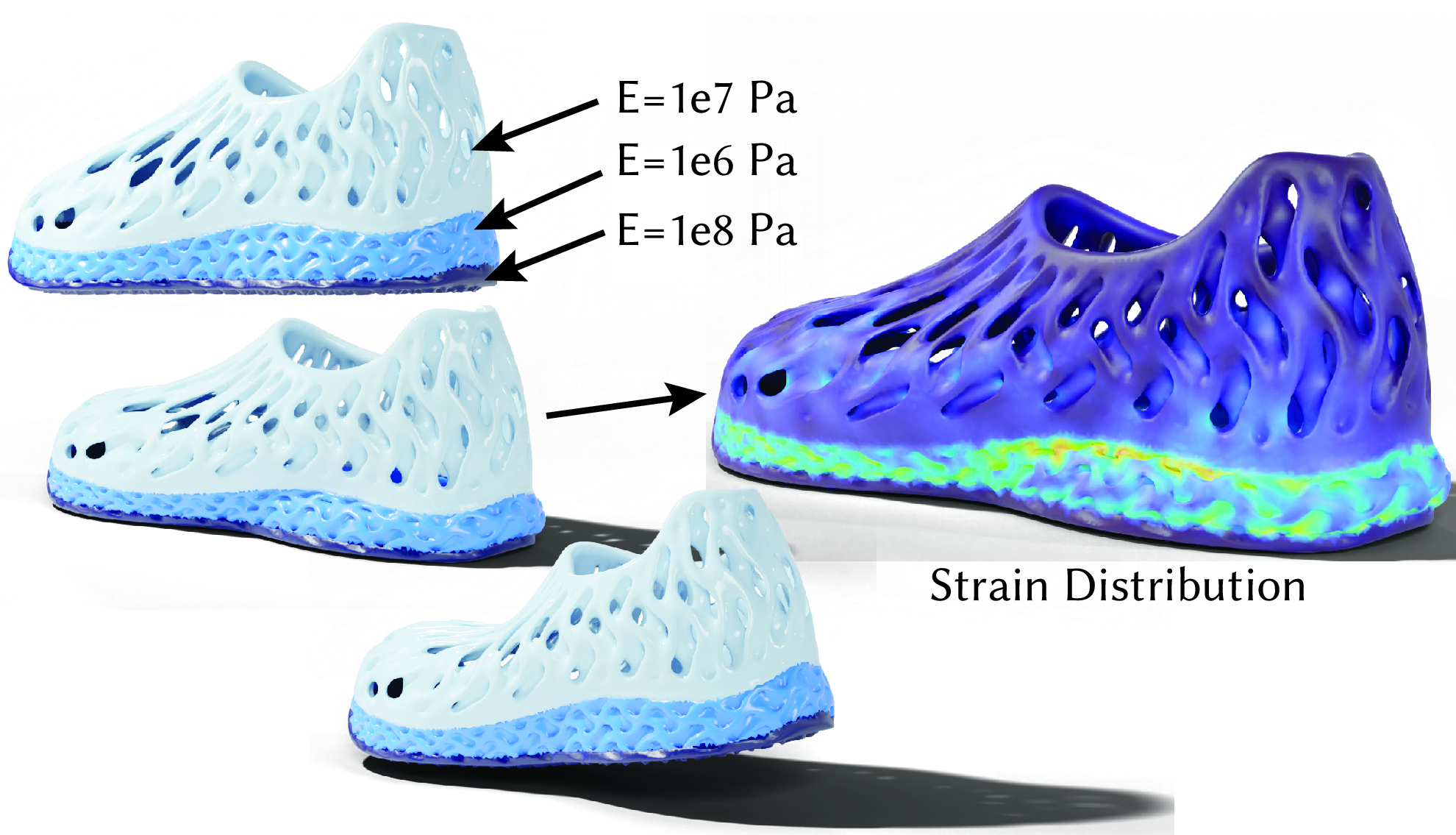}
    \caption{
        Here we drop a shoe with complex geometry and material distribution and evaluate the rebound after impact.
        The shoe is composed of three materials: a soft midsole (E=1e6 Pa), a stiff outer sole (E=1e8 Pa), and a slightly stiff upper (E=1e7 Pa).
        The shoe is dropped at 5/ms onto a plane, at the point of impact we show the strain distribution, highlighting the ability over method to capture the bulging deformation between the stiff components of the shoe.
        With the stored strain energy, the shoe rebounds and we show the resulting deformation at the end of the simulation.
        \ty{feedback on figure. I also can show the partitions and basis functions, since they are interesting. Might warrant a double column figure}
     \label{fig:shoe}}
\end{figure}

\begin{figure}
    \includegraphics[width=\linewidth,keepaspectratio]{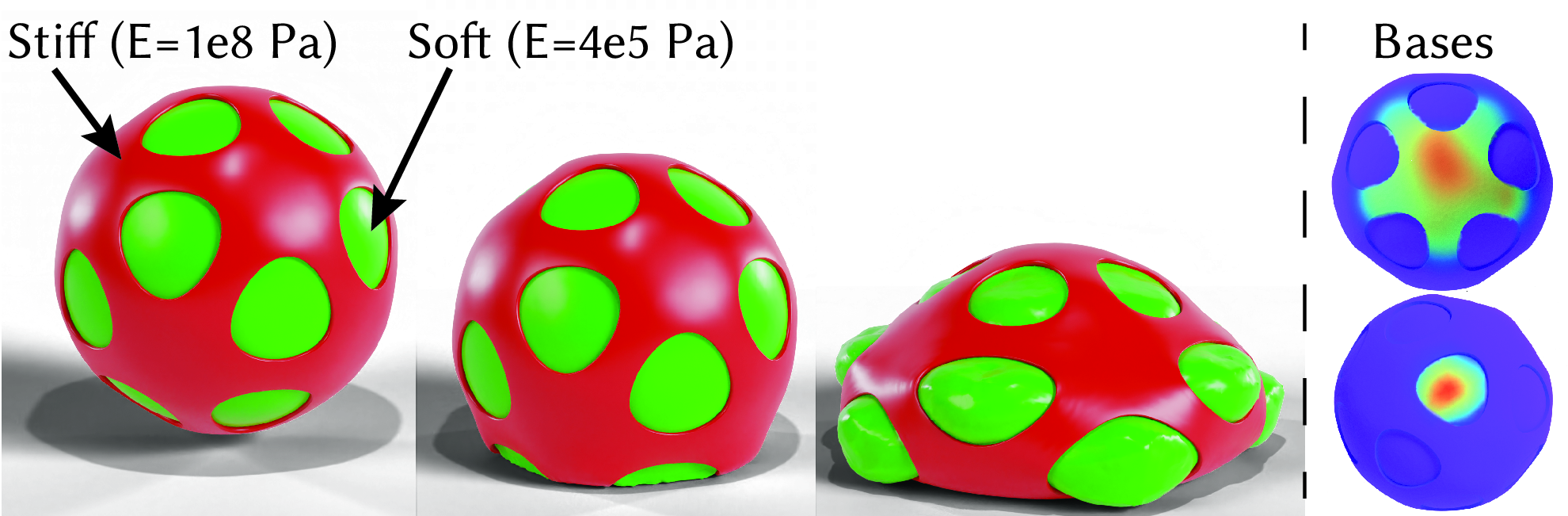}
    \caption{We again highlight the ability of our method to capture complex geometry and material distributions.
    Here we simulate a "stressball" with a stiff (1e8 Pa) outer shell and a soft (4e5 Pa) core at a high speed (10 m/s) impact against the ground.
    The resulting deformation shows the complex bulging within the "cells" of the stress ball, and the lower frequency deformation of the stiff core.
    This is made possible by our material aware subspace bases. On the left we show the partitioning of the domain, showing that we can handle non-convex domains.
    The basis functions defined on them are both material and geometrically conforming, enabling the simulation to capture these complex dynamics.
    \ty{todo: show partitions and basis functions}
     \label{fig:stressball}}
\end{figure}

\begin{figure}
    \includegraphics[width=\linewidth,keepaspectratio]{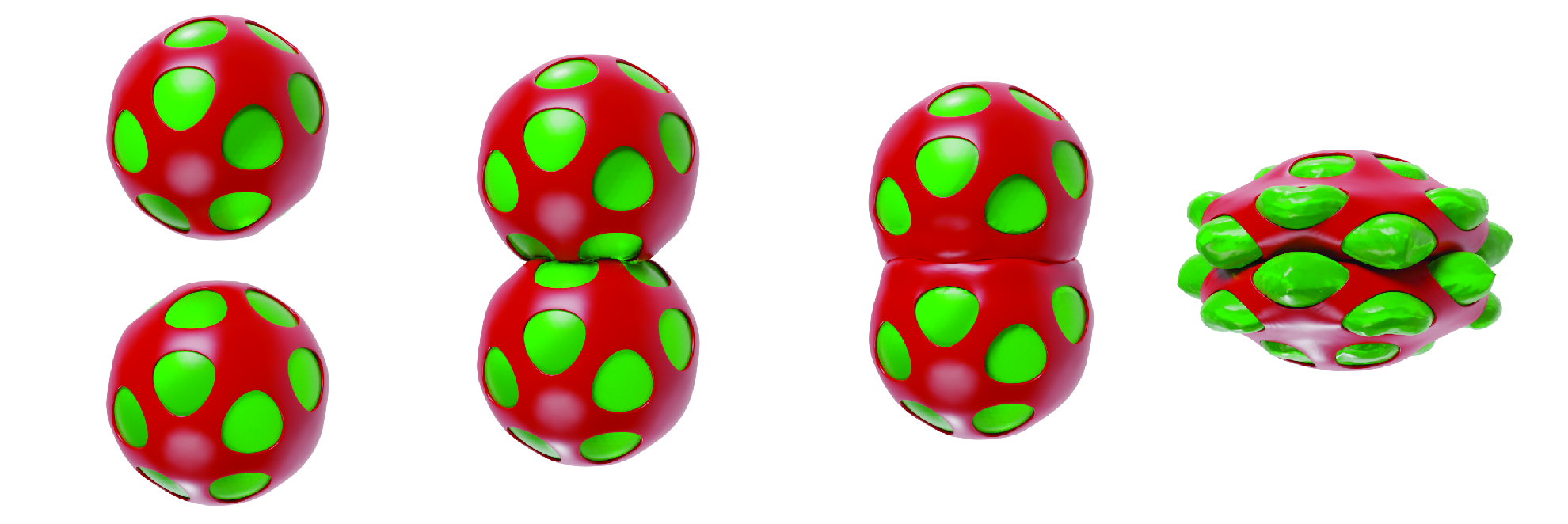}
    \caption{We perform the same simulation but instead of a single stressball, we collide two stressballs at a relative velocity of 20 m/s.
     \label{fig:two-stressball}}
\end{figure}

\begin{figure}
    \includegraphics[width=\linewidth,keepaspectratio]{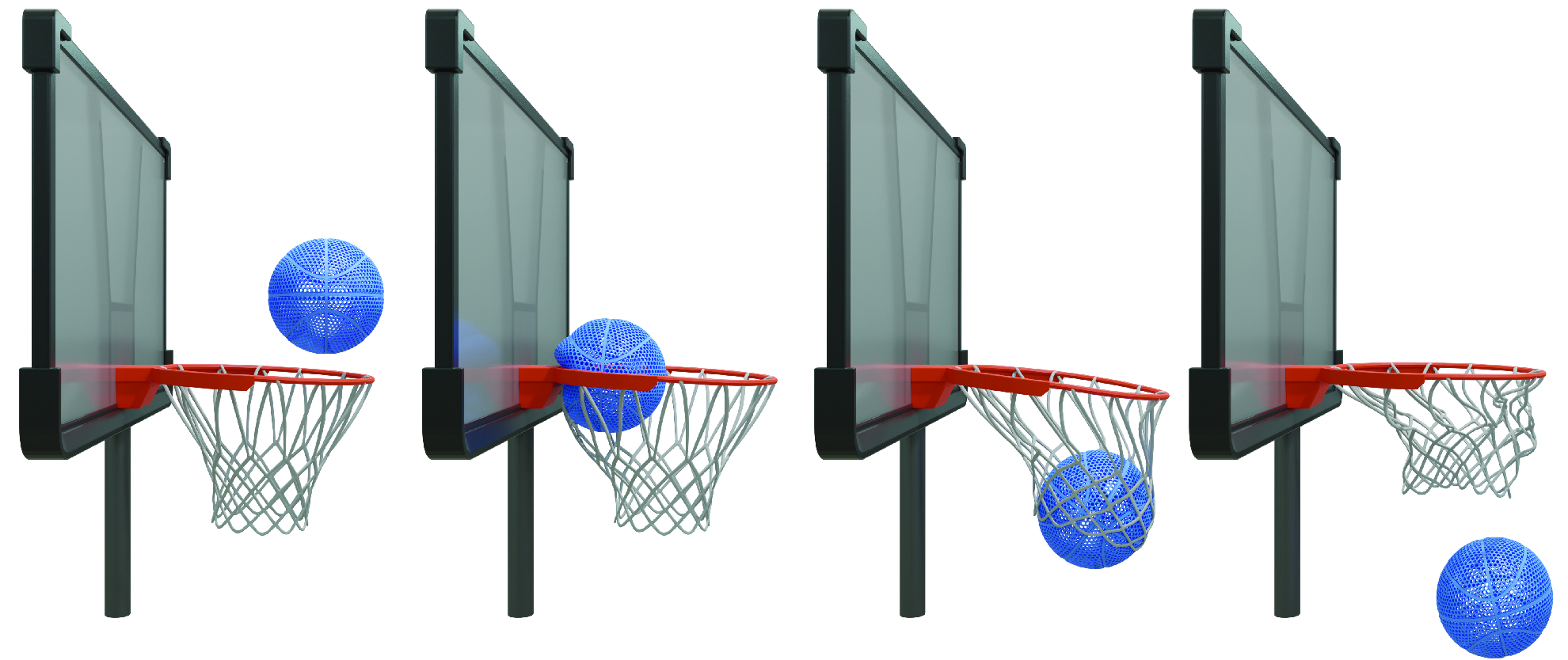}
    \caption{Here we show additional frames from the basketball hoop simulation \reffig{teaser}. We highlight the different types of deformation throughout the simulation: low-frequency stiff motion from the rim, medium-frequency deformation from the basketball impacting the rim, and high-frequency deformation as the ball collides with the net.
     \label{fig:basketball-hoop}}
\end{figure}

\begin{figure}
    \includegraphics[width=\linewidth,keepaspectratio]{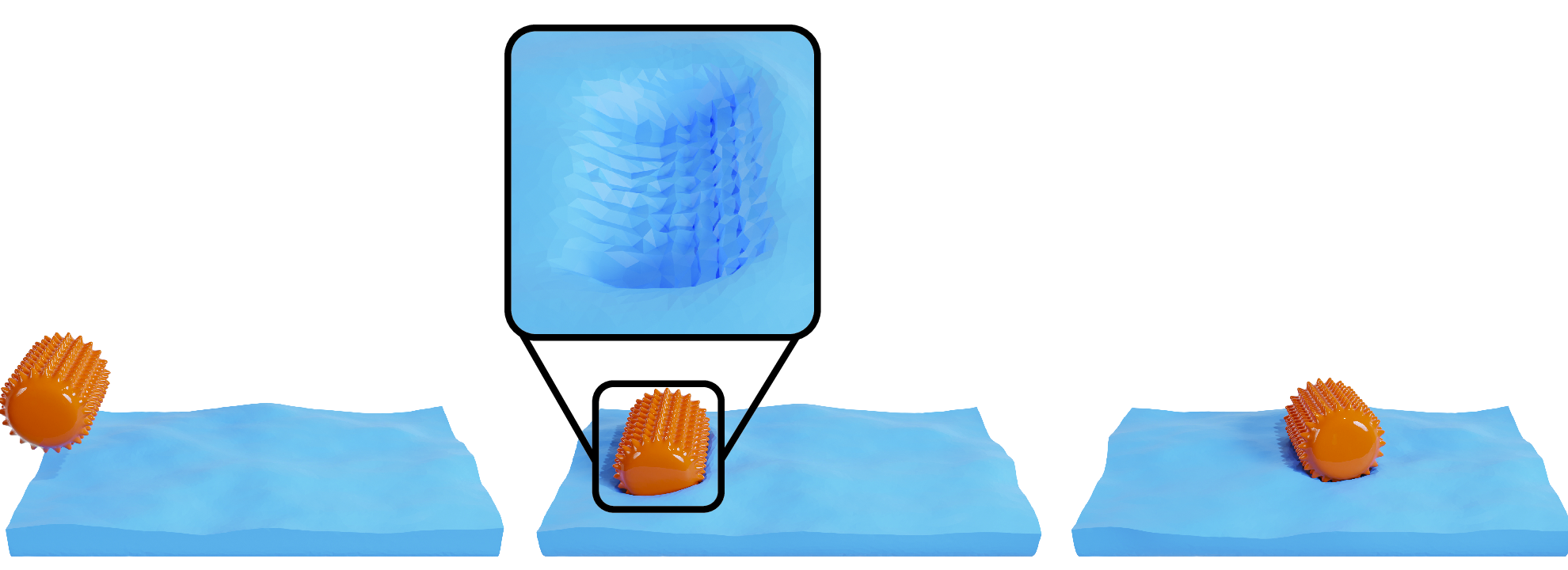}
    \caption{We highlight the handling of Dirichlet boundary conditions with our method on a large scale simulation with a foam roller over a terrain. We recover the high-frequency impacts at the points of contact, as well as the low-frequency oscillations over the soft terrain.
     \label{fig:DBC-roller}}
\end{figure}

\begin{figure}
    \includegraphics[width=\linewidth,keepaspectratio]{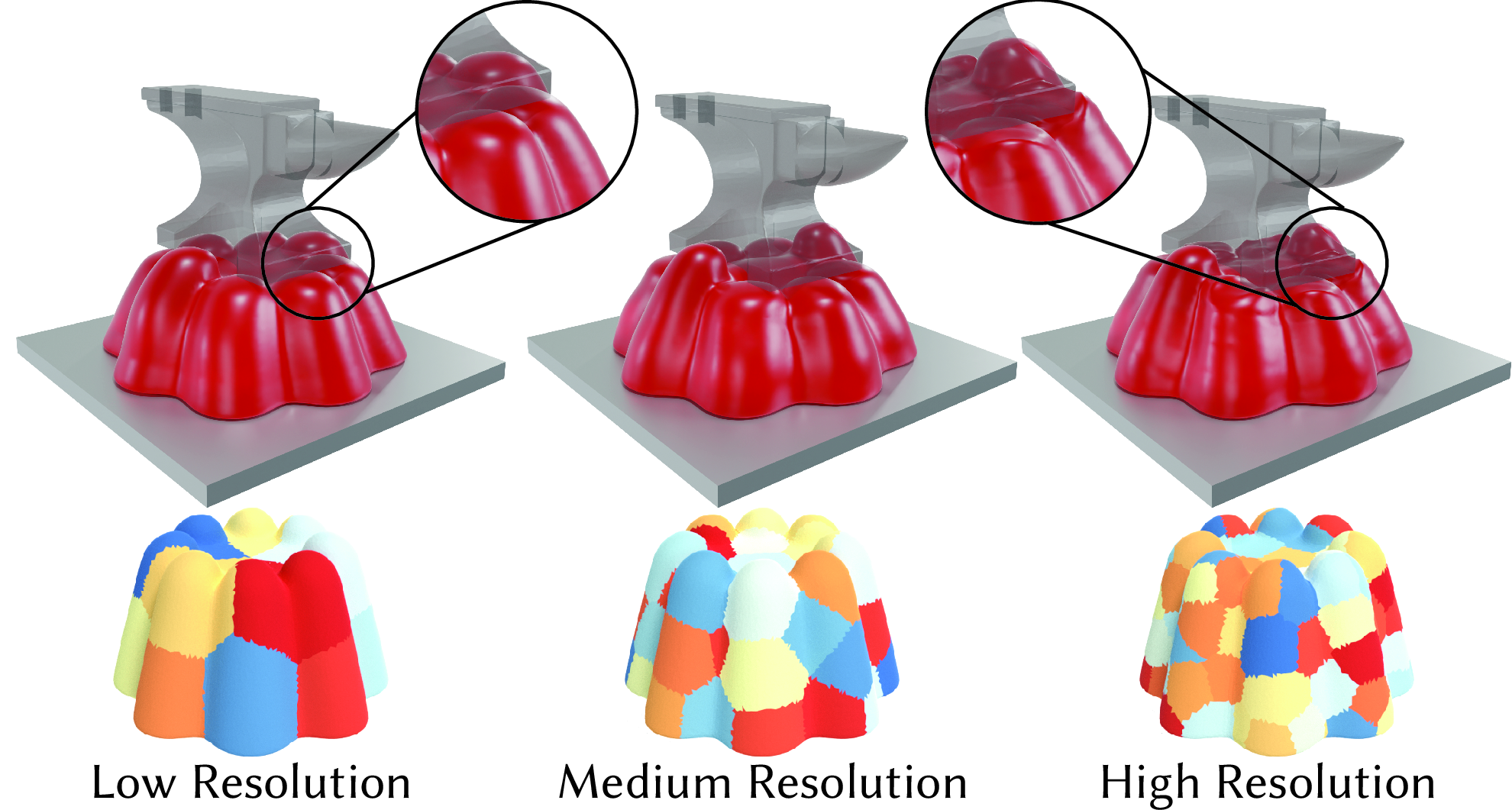}
    \caption{We highlight the convergence of our subspace simulation to the full-space IPC solution. We simulate a jello block with a stiff anvil dropped on it, and vary the subspace resolution (15 handles, 47 handles, and 118 handles). As we increase the resolution we see the subspace recovers more of the full-space dynamics, with the imprint of the anvil becoming more pronounced.
    \ty{Also can mention that the runtime costs are not necessarily faster. The high res variant is actually the fastest among the three of them, with the low res shortly behind it in runtime. The middle resolution is the slowest, taking over 2x longer than the high res.}
     \label{fig:jello-anvil}}
\end{figure}

\paragraph{Complex Geometries and the Power of Refinement}
In \reffig{kooshball} we fire a "kooshball" with complex thin tendril geometries at 5 m/s onto a plane. Comparing output of our method with 10 full-space refinement CG iterations (left), our method with 40 refinement iterations (middle), and the full-space IPC solution (right), we see that the effect of our fine-level CG iterations is dramatic. Here our 40 CG version is visually indistinguishable from the full-space solution, with an 8x speedup. Interestingly we note that the 10 CG version (left) noticeablly lacks the complex interlocking strands, and yet, its total runtime cost is much slower than the 40 CG version. This highlights how our additional full-space work can actually accelerate our total solve, particularly when much of the dynamics is in local deformation.

\paragraph{Heterogenous Distributions}
In Figures\, \ref{fig:teaser}, \ref{fig:shoe}, \ref{fig:stressball}, \ref{fig:two-stressball}, and \ref{fig:basketball-hoop}, we highlight a range of challenging examples efficiently simulated by our method with both complex geometric patternings and material heterogeneities. Importantly to support this our mesh partitioning method is robust to these challenging distributions producing partitions and basis functions that respect and work well with the material boundaries and complex domain shapes.

\paragraph{Boundary Conditions} Boundary conditions on deformable domains can be a challenge to subspace models inequipped to resolve them. As covered, our SMS subspace construction directly incorporates Dirichlet boundary conditions seamlessly, as we see demonstrated in \reffig{DBC-roller}.

\section{Conclusion and Limitations}
\label{sec:future}

We have presented a multi-level elastodynamics timestep solver for accelerating IPC simulations. Our method retains the robustness and generality of IPC, with close-matching visual and per-solve fidelity, while, at the same time offering significant speedup across the board. 
Looking ahead there remain many opportunities for future work to address current limitations, and explore new extensions of this work. 

To begin with, when it comes to our bases, it would be interesting to explore updates to our biharmonic solve (e.g., using a global solve while maintaining compact support regions), provide smoothness guarantees for our weights, and likewise be able to give guarantees of nonzero support within domains, which we cannot guarantee currently. Likewise, further improvements to our Hessian cubature method should be important. In some cases our 
current cubature rule can be insufficient under extreme deformations. In such cases, this leads to underapproximation of stiffness, and so ``overshooting'' in subspace direction, with correspondingly small line-search progress. Here adaptively switching to higher-order rules could be an efficient solution to explore. 

At the same time, looking ahead, we see a number of exciting opportunities in extending our approach. First and foremost, while our current three-level solver is so far highly effective, it does beg the question of how and if our bases and solver strategies can effectively be applied in a more general hierarchical framework (e.g., multigrid). Likewise, we are excited to consider whether our method, likely with additional work in basis construction required, can be extended to the simulation of shell and rod models. Finally, we note that while all stages of our current pipeline, including linear solves, are fully parallelizable, we have so far only explored its application on the CPU. Targeting GPU acceleration with our method brings interesting new challenges\ \cite{haung2024GIPC} that should enable significant additional performance boosts.

\bibliographystyle{ACM-Reference-Format}
\bibliography{main}

\end{document}